\documentclass[preprint2]{aastex631}

\renewcommand{\t}{\theta}

 \shorttitle{Quads from Vanishingly Elliptical Lenses}
\shortauthors{Falor and Schechter}

\graphicspath{{./}{figures/}}

\begin{document}

\title{The Quadruple Image Configurations of Asymptotically Circular Gravitational Lenses}

\correspondingauthor{Paul L. Schechter}
\email{schech@.mit.edu}

\author[0000-0002-0511-1848]{Chirag Falor}
\affiliation{MIT Department of Physics, \\
Cambridge, MA 02139, USA}

\author[0000-0002-5665-4172]{Paul L. Schechter}
\affiliation{MIT Department of Physics, \\
Cambridge, MA 02139, USA}
\affiliation{MIT Kavli Institute for Astrophysics and Space Research, \\
Cambridge, MA 02139, USA}








\begin{abstract}

The quadruple image configurations of gravitational lenses with
vanishing ellipticity are examined.  Even though such lenses
asymptotically approach circularity, the configurations are stable if
the position of the source {\it relative} to the vanishing diamond
caustic is held constant.
The configurations are the solutions of a
quartic equation, an ``Asymptotically Circular Lens Equation" (ACLE),
parameterized by a {\it single} complex quantity.  Several alternative
parameterizations are examined.  Relative magnifications of the images
are derived.  When a non-vanishing quadrupole, in the form of an
external shear (XS), is added to the singular isothermal sphere (SIS),
its configurations emerge naturally as stretched and squeezed versions
of the circular configurations.  And as the SIS+XS model is a good
first approximation for most quadruply lensed quasars, their
configurations likewise have only $2+1$ salient dimensions. The
asymptotically circular configurations can easily be adapted to the
problem of Solar System ``occultation flashes.''
\end{abstract}




\section{Introduction} \label{sec:intro}

The simplest models for quadruply lensed quasars (henceforth quads)
require seven parameters.  Four of these are relatively uninteresting,
governing position, angular orientation on the sky, and an overall
scale.  The interesting variations among different quads are due to
the three remaining degrees of freedom.  These three are
straightforwardly parameterized by the $x$ and $y$ coordinates of the
displacement of the source from the line of sight to the center of the
potential and the amplitude of the potential's quadrupole term.

In the present paper, we further simplify the problem by analyzing the
quads formed by an asymptotically circular lens, reducing the number
of salient parameters to two. This two dimensional model space can be
described solely by the relative position of the source with respect
to the center of the potential.

Lens experts might reflexively object that, as the
ellipticity of a potential approaches zero its ``diamond caustic"
-- inside which the source must lie to produce four images 
\citep{Ohanian_1983} -- must also vanish. But if we correspondingly
scale the distance of the source from the center of the lens, keeping
its position {\it relative} to the diamond caustic constant, we get a stable
configuration of four images on a circle.

The quadruple configurations of the vanishingly elliptical lens can be
(and have been) found in a variety of seemingly different contexts.
But the connections among these multiple resurfacings have been hidden
by different parameterizations, different labeling conventions
and different emphases.

Here we describe
three distinct parameterizations for these configurations,
two of which are in terms of model parameters and one of which
is in terms of observable quantities.

We begin, in Section 2, with the Witt-Wynne geometric solution of the lens
equation for the singular isothermal elliptical potential (SIEP)
of \cite{Schechter_2019}, not because of chronological precedence but
because it displays the solutions graphically. 

The circular Witt-Wynne construction can be solved algebraically,
yielding a quartic equation for the polar angles of the four images.
We call this the ``Asymptotically Circular Lens Equation'' (henceforth
ACLE).  It reappears in subsequent sections and serves to establish
the connection between superficially different circumstances.
\cite{WilliamsWoldesenbet}
give closed form solutions for the
four images in their discussion of the singular isothermal quadrupole
potential (henceforth SIQP).

This quartic equation was first derived
by
\cite{KassiolaKovner} in their
discussion of the lens equation for the SIQP, which we review in Section
3.
Their polar angles, measured with respect to the center of the
lens potential, are independent of the strength of the quadrupole, but
they do not explicitly address the asymptotically circular case.
Kassiola and Kovner isolated a relation among the four angles, their
``configuration invariant", that in effect establishes the two
dimensionality of the solution subspace.

For both the vanishingly elliptical SIEP and the SIQP, the space for
which the ACLE yields four distinct solutions is bounded by an
astroid.  In Section 4 we introduce a new coordinate system in which the radial
distance from the origin is measured as a fraction of the distance to
that astroid.

The parameterizations of the quadruple configurations of
asymptotically circular lenses are not limited to using model
quantities.  \cite{WilliamsWoldesenbet}  showed that the
angular solution space of the SIQP could be completely described in
terms of two differences of the observed angular coordinates of its images.
This yielded a two dimensional ``Fundamental Surface of Quads" (henceforth FSQ).
We derive it from the ACLE in Section 5.  The great majority of quadruply
lensed quasars lie close to this surface despite the fact that their
potentials are non-circular and better fit by models other than the
SIQP.

In Section 6 we reintroduce the third dimension by examining the singular
isothermal sphere with external shear (SIS+XS).  We show that for the
same relative source position within the diamond caustic, the
configurations with shear are simply ``scronched''\footnote{We use the
word ``scronch", adopted by \cite{ellenberg2021shape} in his popular
book {\it Shape}, to describe stretching in one direction and squeezing
in the orthogonal direction.} versions of the circular
configurations.  In Section 7, we revisit the SIEP allowing for
non-vanishing ellipticity.

Though he does not explore image configurations, \cite{An_2005} shows
that the two dimensional solution space of the ACLE is not restricted
to isothermal potentials and applies more generally to non-isothermal
potentials with vanishing quadrupoles.  
\cite{Saha_2003} had previously noted that
{\par\narrower{
``...properties that arise because of the breaking of circular
symmetry are relatively model independent and robustly reproduced by
even a rudimentary model."
\par}
\par
}
\smallskip
In Section 8 we arrive at the same conclusion by an alternative and
perhaps more straightforward route, considering the case of
circularly symmetric potentials with vanishing external shear.

The ACLE is of use even beyond gravitational lensing. In Section 9 we
consider the quadruple image configurations calculated for a ``central
flash" observed during a 1989 occultation by Saturn \citep{Nicholson_1995}.
We find they lie closer to the FSQ than do the observed
configurations of known quadruply lensed quasars.

In Section 10 we discuss the relative flux ratios for the ACLE quadruple image
configurations. 

\section{Quadruple Configurations of the Asymptotically Circular Lens from
  the Witt-Wynne construction}\label{sec:ACP_geo}

\begin{figure*}
\gridline{\fig{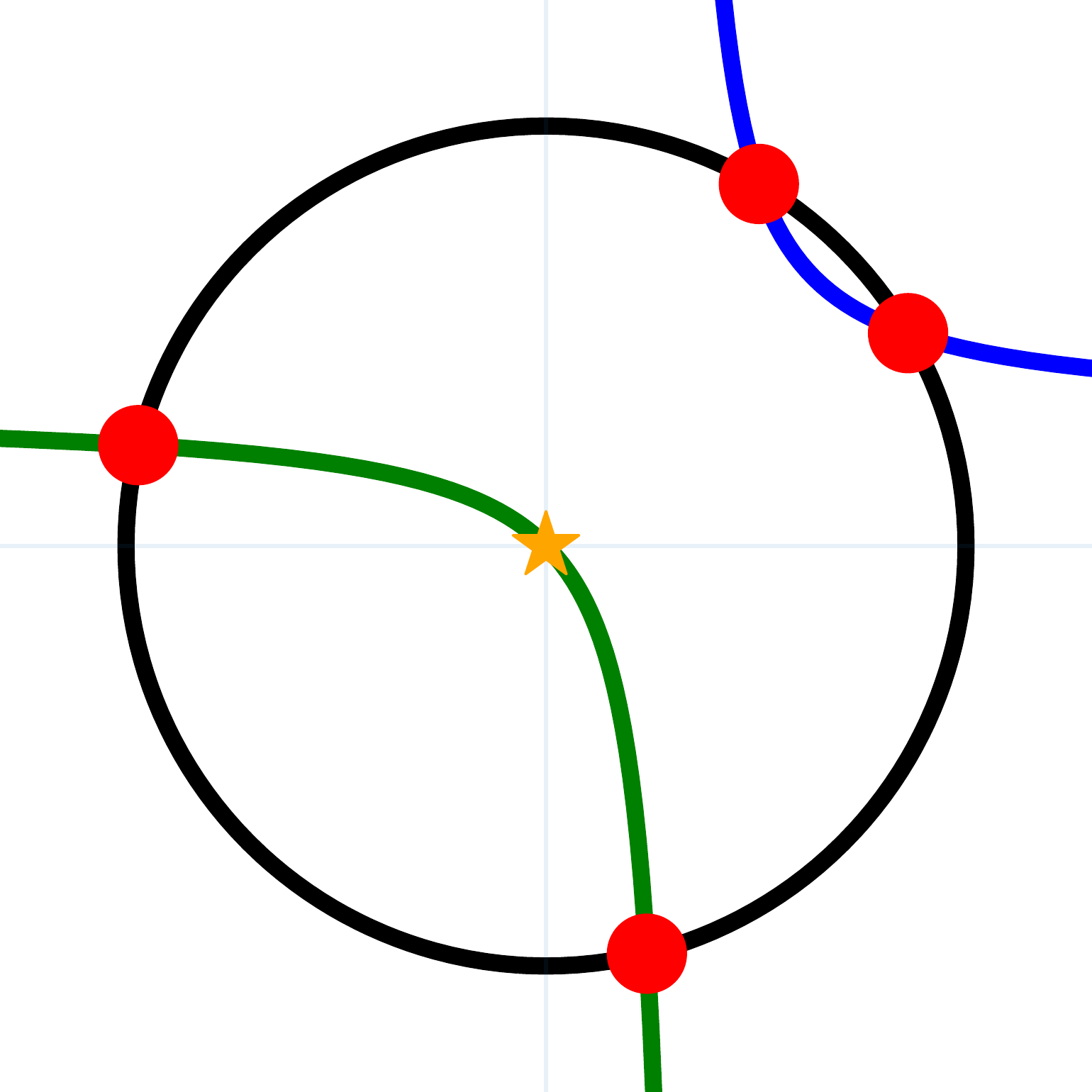}{0.3\textwidth}{(a)}
          \fig{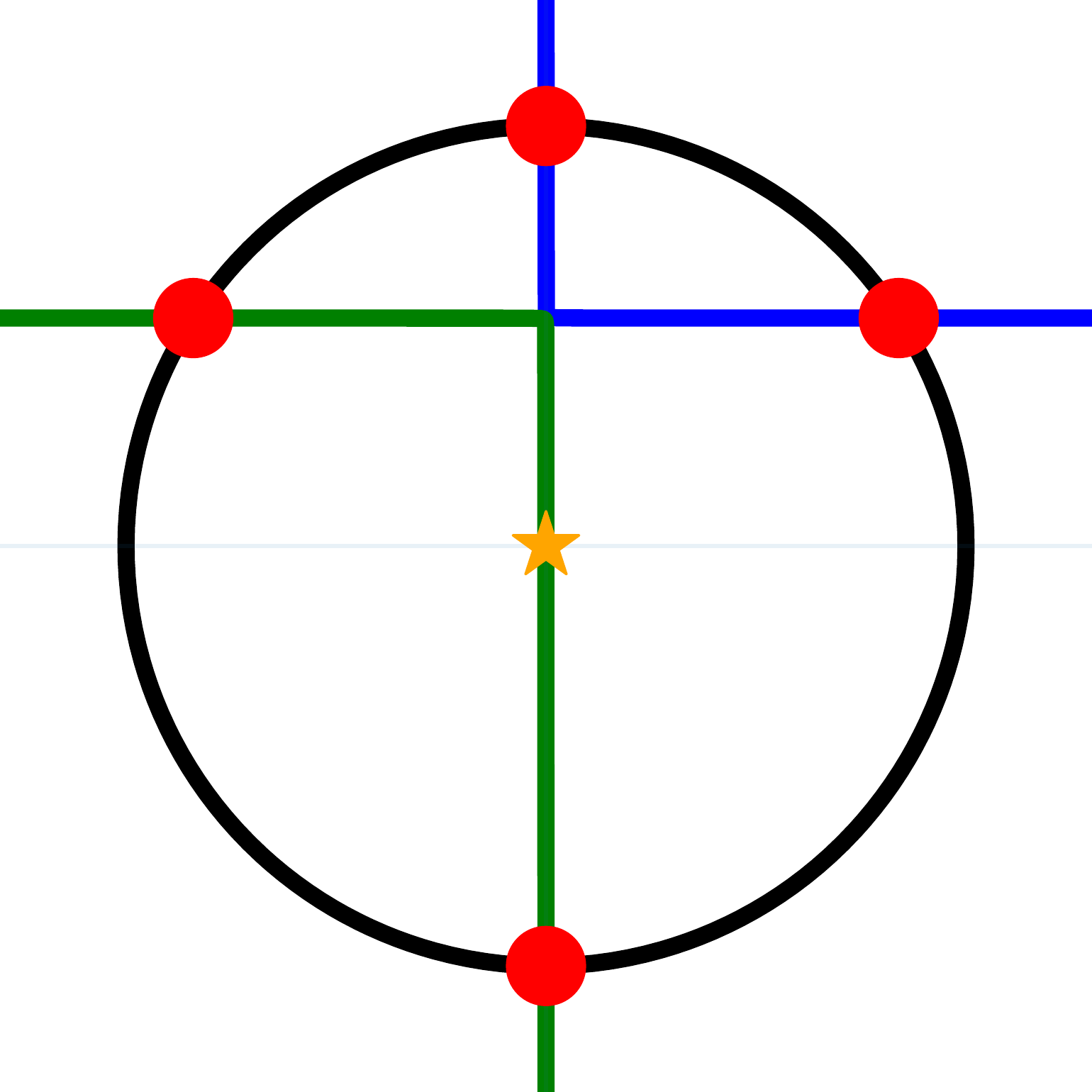}{0.3\textwidth}{(b)}
          \fig{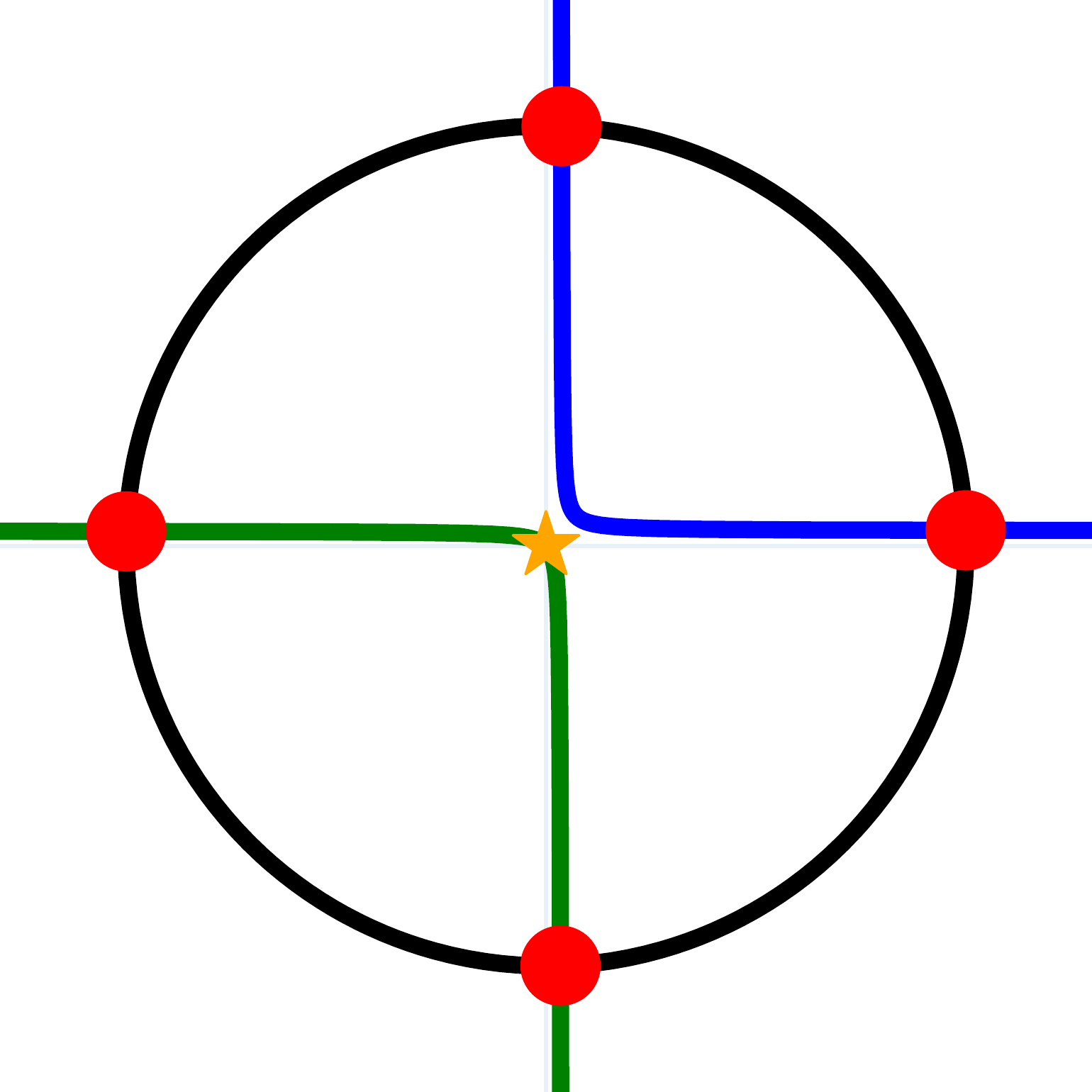}{0.3\textwidth}{(c)}}
\caption{The plots show the construction of several asymptotically circular image configurations.  Panel (b) shows a ``kite'' configuration,
symmetric around one of the two diagonals connecting the images.
The system in panel (c) can be thought of as a ``super-kite'',
symmetric around both diagonals.
The system in panel (a) is symmetric about neither diagonal,
and hence an ``un-kite.''
\label{fig:Witt-Wynne construction}}
\end{figure*}

\cite{Schechter_2019} present a simple geometric construction
that gives 4 images for an SIEP.  Their `recipe' proceeds as follows:
{\par\narrower
{``(1) Find the rectangular hyperbola that passes through the points. (2) Find the aligned ellipse that also passes through them. (3) Find the hyperbola with asymptotes parallel to those of the first that passes through the center of the ellipse and the 
pair of images closest to each other.''
\par}
\par}

The center of the ellipse gives the position of the source, and the
asymptotes of the hyperbola are aligned with the major and minor axes
of the potential.  The ellipse has the same axis ratio as the
potential, but lies perpendicular to it.

Nothing in the above recipe requires that the ellipse have non-zero
ellipticity.  For an SIEP of vanishing ellipticity, Wynne's ellipse
becomes a circle.  This suggests that by inverting the Witt-Wynne
construction, we can use its limiting cases to produce the quadruple
configurations formed by asymptotically circular isothermal
potentials.

\subsection{Edge case of the Witt-Wynne recipe}
We invert the \cite{Schechter_2019} recipe as follows:

\begin{enumerate}

\item Start with a unit circle, the 
 limiting case of Wynne's ellipse \citep{wynne2018robust}.
 
\item  Draw a rectangular hyperbola with arbitrary
semi-major axis,
whose asymptotes (with no loss of generality) are parallel to the $x$-
and $y$-axes and one of whose branches passes through the center of
the circle.  This is our Witt hyperbola \citep{Witt_1996}.

\item The points of  intersections of these two conic sections
give the positions of the images of a quadruply lensed quasar.\footnote{Some
combinations of semi-major axis
for the hyperbola and position of
the center of the circle on
that hyperbola give
only 2 intersection points. These are valid double lensed quasar
configurations, but we focus here on the quads.}
\end{enumerate}


The $x$ and $y$ axes are the symmetry axes of the SIEP by virtue of
having constructed the asymptotes of the hyperbola parallel to them.


Figure
\ref{fig:Witt-Wynne construction}
shows some examples with the
associated circle and hyperbola for three different configurations. To
qualitatively describe the quads, we use the classification system
of \cite{Saha_2003}. Figure
\ref{fig:Witt-Wynne construction}(a)
is
an example of the ``inclined quad" configuration, where the source lies
on  neither axis of the diamond caustic.  Figure
\ref{fig:Witt-Wynne construction}(b),
where the source lies on
either axis of the diamond caustic, would form a ``long" or ``short
axis quad", if the potential were not circular. Figure
1(c) shows a ``core" configuration,
where the source lies almost at the center of the diamond caustic.  The center of the potential is displaced infinitesimally from the
source.

\subsection{Two-dimensionality of the asymptotically circular model space of Witt-Wynne configurations}

Recall that we are not concerned about the scale, rotation, or absolute
position of configurations, so we can neglect the orientation and
position of the chosen random rectangular hyperbola and the size
of the circle. The
only two degrees of freedom are the ratio of the major axis of the
hyperbola to that of the radius of the circle and the position of the
center of the circle on the rectangular hyperbola, characterized by
one parameter each.  Hence, the configurations generated in the
circular limit of Witt-Wynne method are two-dimensional.

\subsection{The Asymptotically Circular Lens Equation}\label{subsec: ACP lens eqn}
The image configurations of a circular potential are described by the four angles $\theta$ around the circle. In Appendix
\ref{app:ACPACE}
we show that for Witt's hyperbola centered at $(x_h, y_h)$ 
and a Wynne circle of radius $b$ centered at the origin, the four angles are the solution of
\begin{equation}\label{eqn:Main_ACP_equation}
    e^{4i(\t-\psi)} - 2W e^{3i(\t-\psi)} + 2\overline{W}e^{i(\t-\psi)} - 1 = 0,
\end{equation}
where
\begin{equation}\label{eqn:WforSIEP}
    W = \frac{x_h + iy_h}{b}.
\end{equation}
Equation (\ref{eqn:Main_ACP_equation}) has exactly the same form, up to
multiplicative constants, as equation (2.4) of Kassiola and Kovner,
who derived it for solutions of the SIQP potential with finite
quadrupole, which we discuss in Section 3.
But it is applicable far beyond, so we refer to it as the ``Asymptotically
Circular Lens Equation'' (ACLE).  We show in Section 8 that the ACLE, while derived
using singular isothermal potentials,  gives solutions for all
nearly circular potentials with vanishing quadrupole.

The parameter $\psi$ (the position angle of one of the
two symmetry axes) merely rotates the configuration. The roots of the
ACLE,
$\t_1, \t_2, \t_3,$ and $\t_4$, are the angular positions of the four images. Applying Vieta's formula, we have
\begin{equation}\label{eqn:psi_def}
    \psi = \frac{\t_1 + \t_2 + \t_3+ \t_4}{4} \pm \frac{\pi}{4}
\end{equation}
a result that also traces back at least as far as \cite{KassiolaKovner}.

\cite{WilliamsWoldesenbet} give closed form expressions\footnote{It appears that a factor of $6(2^{2/3})$ in these
expressions was incorrectly transcribed as $(62)^{2/3}$.}
for the four
solutions to the ACLE.  In Appendix
\ref{app:closedform}
we recast
the closed form solution retaining the original
complex parameters $W$ and $\overline{W}$.  The four solutions are 
given by a single equation,

\begin{eqnarray}
 2 e^{i\theta} & = &  W \widehat{\pm} \sqrt{u+W^2} \label{eqn:explicit}\\
  &\widetilde{\pm}& \sqrt{(W \widehat{\pm} \sqrt{u + W^2})^2 - 2\left(u \widehat{\pm} \sqrt{u^2 + 4}\right)  \nonumber }
\end{eqnarray}
with a solution
for each pair of choices of $\widehat{\pm}$ 
and $\widetilde{\pm}$,
and where complex $u$ is the solution to the cubic
\begin{equation}\label{eqn:cubic}
   u^3 + 4(1-W\overline{W})u +4(W^2-\overline{W}^2) = 0 
\end{equation}  

The solutions to the ACLE are governed entirely by a single complex
number.  The space of quadruple configurations for lenses of
vanishing ellipticity is again seen to be two dimensional.

We show in Appendix
\ref{app:ACPACE}
that if
Equation (\ref{eqn:Main_ACP_equation}) has four distinct solutions for
$\theta$, $W$ must lie inside a unit astroid.  Adopting a coordinate
system with $\psi = 0$, the astroid is aligned with its axes,
\begin{eqnarray}
    W = p + iq, \qquad p, q \in \mathbb{R} \label{eqn:wdefinition} \\
    p^{2/3} + q^{2/3} < 1. \label{eqn:boundary4}
\end{eqnarray}
In this aligned coordinate system, the ACLE takes on the deceptively
simple form
\begin{equation}\label{eqn:deceptive}
  p \sec \theta + q \csc \theta = 1 \quad .
\end{equation}
\begin{figure}[ht]
    \centering
    \includegraphics[width=0.4\textwidth]{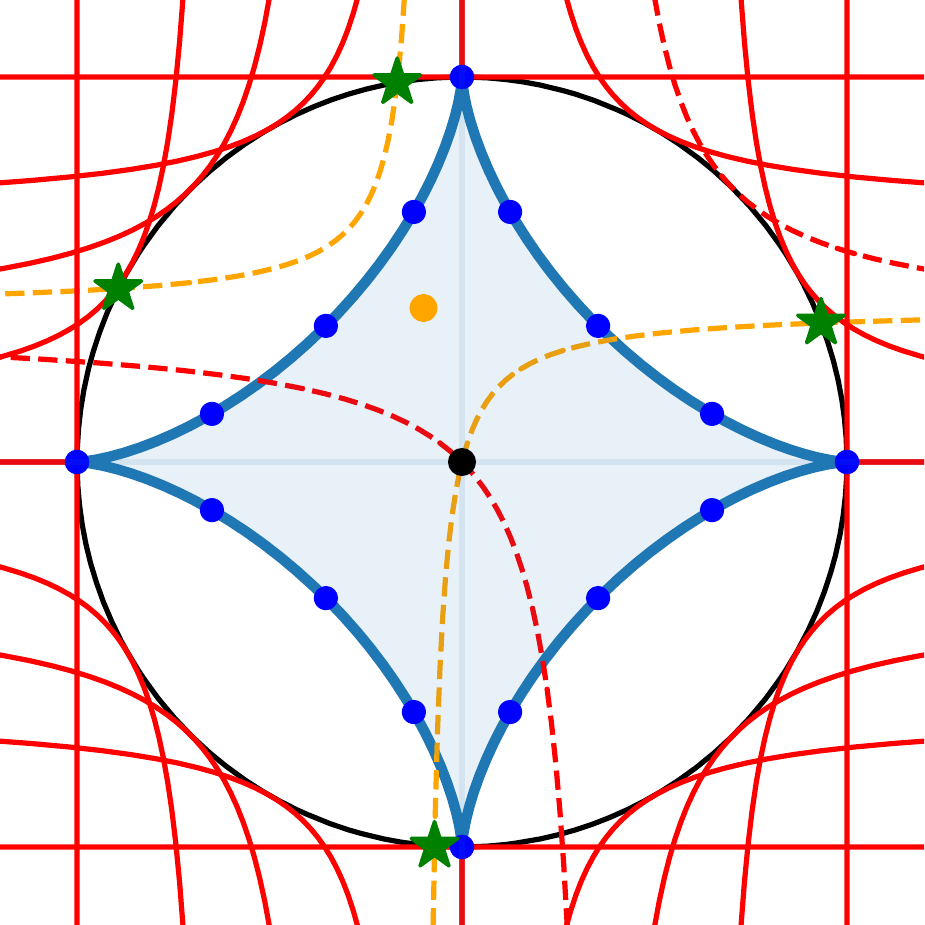}
    \caption{The blue dots are the centers of the Witt's hyperbolae (red), which are drawn tangent to the Wynne's circle (black). A configuration (green stars) is constructed for an arbitrary position of center of Witt's hyperbola (yellow) to illustrate how configurations are generated from points inside the Witt-Wynne diamond.}
    \label{fig:tangent_hyperbola_astroid}
\end{figure}
\subsection{Hyperbolae tangent to the Wynne circle and the astroid formed by their centers}\label{subsec: hyperbola, witt-wynne diamond}
In the Witt-Wynne construction, the image configurations are
determined solely by the position of the center of Witt's hyperbola,
$W$.  The hyperbola is constrained to pass through the center of
Wynne's circle (which coincides with the source position).  One
can construct the astroid that bounds the four image solutions by
plotting the centers of the hyperbolae tangential to the circle.

Two images (e)merge and (dis)appear at the point of tangency.  The
centers of the tangent hyperbolae demarcate the envelope of the four image
region.  Figure
\ref{fig:tangent_hyperbola_astroid}
shows how
these centers trace the astroid inside which Witt's hyperbola gives
four images.

Using this same construction for an SIEP with finite ellipticity, one
obtains a scronched astroid, the discussion of which we defer
until Section 6.  In Section 5 we discuss the Witt-Wynne
construction for SIS+XS potentials, which likewise give
scronched astroids.

The ``Witt-Wynne diamonds,'' traced by the center of Wynne's
hyperbola, should {\it not} be mistaken for Ohanian's (1983) diamond
caustics, traced by the source position (although they are clearly
related).  In general the diameters of those diamond caustics
are not equal.  For both the SIEP and an SIS+XS, the diamond caustic
is elongated perpendicular to the elongation of the Witt-Wynne
diamond.  The center of the hyperbola and the source lie in different
quadrants of the complex plane.  And for the case of the SIEP, the
diamond caustic differs, albeit subtly, from a scronched astroid.

\subsection{
The center of Witt's hyperbola from the average image position}
\label{subsec:hyp_centroid}
The center of Witt's hyperbola can be found by expressing it as a conic section,
with four unknown coefficients, evaluating it at each of
the four images and solving for coefficients by inverting a
$4\times4$ matrix. Alternatively, we show
in Appendix A that  $(x_h, y_h)$, the coordinates of the
center of Witt's hyperbola relative to the center of Wynne's
circle can be calculated from the coordinates of the centroid
of the four images and are given by $2(x_{\textrm{\scriptsize
centroid}},y_{\textrm{\scriptsize centroid}})$.

\section{The ACLE derived from
the not-at-all circular SIQP}
\label{sec:ACP_alg}
In the previous section, we
constructed geometrically
the angular configurations generated by an
asymptotically circular singular isothermal potential.
Identical {\it angular}
configurations arise for the
singular isothermal quadrupole potential, even when the amplitude
of the  quadrupole is far from vanishing.  The configurations differ,
however, in that the {\it radial} deflections vary with angle.
The properties of the SIQP have been extensively studied in the literature 
\citep{Kochanek_1991}, \citep{ KassiolaKovner}, \citep{Kormann_1994}
and \citep{Dalal_1998}, but without taking note of the limiting
circular case.

\subsection{Configurations of the SIQP and equivalence to those
  of the ACLE}\label{subsec:SIQP to ACP}
We consider the singular isothermal quadrupole potential (SIQP),
\begin{equation}
    \Phi(r,\t) = br[1-\epsilon \cos{ 2(\t - \psi)}]
\end{equation}
Here $\psi$ is the position angle of the major axis of the quadrupole
of the SIQP and $\epsilon$ is proportional to the quadrupole
moment\footnote{Our sign convention for the quadrupole moment is chosen so
that equipotentials are stretched in horizontal direction for positive
values of $\epsilon$. }.

Starting with the angular component of the lens equation
\citet{KassiolaKovner} derived an equation for the image
positions that is identical, modulo multiplicative constants,
to our Equation (\ref{eqn:Main_ACP_equation}), the ACLE.
Treating the position of the source relative to the
potential as a complex number
\begin{equation}\label{eqn:sourcepos}
  s=re^{i(\phi_s-\psi)} \quad ,
\end{equation}
taking $\psi$ to be  the position angle of the major axis
and $\theta$ to be the angular position of an image measured from
the center of the lens.  
\citet{KassiolaKovner}'s
polynomial equation (2.4) for the positions of the images can be
rewritten as
\begin{equation}\label{eqn:original_configurational_equation}
    e^{4i(\theta-\psi)} - \frac{\bar{s}}{2\epsilon b}e^{3i(\t-\psi)} + \frac{s}{2\epsilon b}e^{i(\t-\psi)} - 1= 0.
\end{equation}
This becomes identical to Equation (\ref{eqn:Main_ACP_equation}), the asymptotically circular lens equation,  if we take
\begin{equation}\label{eqn:WforSIQP}
    W=\frac{\bar{s}}{4\epsilon b}\quad.
\end{equation}

\emph{All} finite amplitudes of the quadrupole in the range $ -\frac{1}{5} \leq \epsilon \leq \frac{1}{5}$ \citep{Finch_2002},
including {\it a fortiori} the asymptotically circular case when
$\epsilon \rightarrow 0$,  give the same set of four image angular
configurations.

It is reassuring that angular configurations for the asymptotically
circular SIQP are the same as those for the asymptotically circular
SIEP of Section 2.  But note that for the SIEP, the complex
quantity $W$ in the ACLE is defined, in Equation (\ref{eqn:WforSIEP}), in terms of the
position of the center of Witt's hyperbola, while for the SIQP, it is
defined in terms of the position of the source.  As with the SIEP,
the SIQP gives four images if $W$ lies within a unit astroid.

\citet[eq. 49]{Kormann_1994}\footnote{$k$ and $a/k$ in their potential correspond to our $b$ and $\epsilon$ respectively.}, tell us that for a potential
aligned with one of the axes,  the diamond caustic of the SIQP has coordinates $(x_a, y_a)$ given by 
\begin{eqnarray}\label{eqn:astroid_parametric_SIQP}
    x_a = ~4\epsilon b \cos^3 \t_c,\nonumber \\
    y_a = -4 \epsilon b \sin^3 \t_c.
\end{eqnarray}
where $\t_c$ is the angular position at which two images merge on the
critical curve.  For the SIQP the diamond caustic is therefore a true
astroid\footnote{
\citet{Ohanian_1983} appears to be the first to have
used the word ``astroid" to describe the diamond caustic.
We reserve the word for true astroids, and use ``astroidal''
for ``scronched'' astroids.  \cite{Chang_1979} show an astroid-like caustic but
do not name it.
} with the equation,
\begin{equation}\label{eqn:SIQP_astroid}
    \left(\frac{x_a}{4 \epsilon b}\right)^{2/3} + \left(\frac{y_a}{4 \epsilon b}\right)^{2/3} = 1.
\end{equation}
But since four image solutions to the ACLE must lie within
an astroid, we might also have arrived at this result using
equations
(\ref{eqn:wdefinition}) and (\ref{eqn:boundary4}).
The angular image configurations for the SIQP, therefore depend
only upon the position of the source relative to the center of the
potential, $s$, scaled by the size of the astroidal caustic. 
In  Section 4, we explore how the configurations vary as the source position varies within the astroid.

\subsection{The Kassiola-Kovner Invariant}
The coefficient of the $e^{2i\theta}$ term in the ACLE vanishes. We show in Appendix
\ref{biquadratic_2_0}
that if $\t_1, \t_2, \t_3,$ and $\t_4$ are distinct roots of the ACLE, then the following identity holds
\begin{eqnarray}\label{eqn:KKinvarianteq}
    \cos\frac{\theta_1 + \theta_2 - \theta_3 - \theta_4}{2} + \cos\frac{\theta_1 + \theta_3 - \theta_2 - \theta_4}{2} \nonumber \\+ \cos\frac{\theta_1 + \theta_4 - \theta_2 - \theta_3}{2} = 0.\quad \quad
\end{eqnarray}
This invariant of the SIQP was discovered by \cite{KassiolaKovner}, who called it the ``configuration invariant''.
But beyond the SIQP, the Kassiola and Kovner invariant is exactly $0$ for {\it any} lens satisfying the ACLE.
Moreover, its invariance holds approximately for many  known lenses \citep{KassiolaKovner}. In
Section 4 we explore its connection to the ``Fundamental Surface of Quads'' (FSQ), found independently by \citet{WilliamsWoldesenbet}.

\section{Parameterizing asymptotically circular lens configurations
  with observed angular differences}
\label{FSQ and KK plot}

In the preceding sections we use the {\it model} parameters of the
asymptotically circular lens to describe the two dimensional space of
quadruple image configurations.  Here we examine the configurations'
three observed angular differences, which  Woldesenbet and Williams
have shown lie on a two dimensional surface, the FSQ.
Any two of the observed differences can be used to derive the third.

\subsection{Labeling convention}\label{subsec:Labeling_convention}

\begin{figure}[ht]
    \centering
    \includegraphics[width=0.4\textwidth]{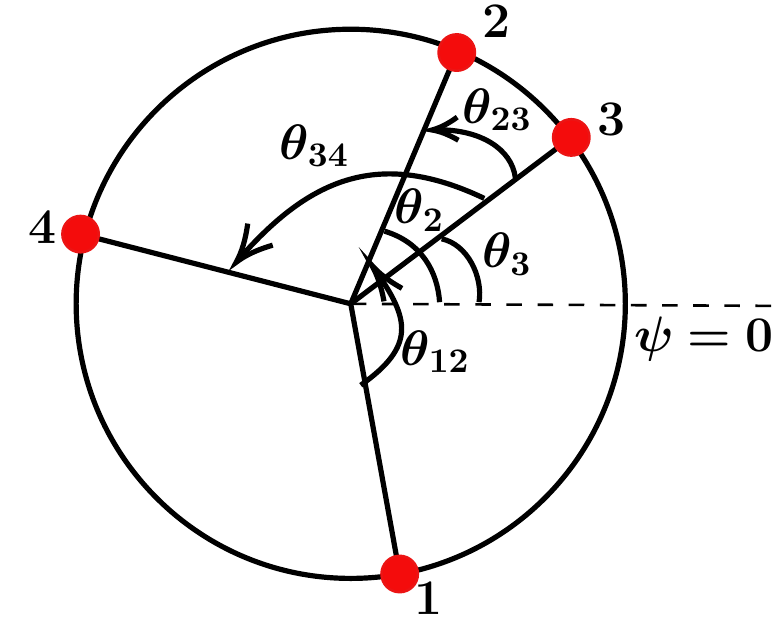}
    \caption{For consistency with the notation of
      \citet{WilliamsWoldesenbet}, we label the images as shown in
      this figure.  Images $2$ and $3$ are the closest pair of images.
      The quadrant in which they lie depends upon
      the angular position of the source relative to the lens,
      $\phi_s$.
      For the configuration shown,       
      $\phi_s$ lies in the $4^\textrm{\scriptsize th}$ quadrant}
    \label{fig:labeling_convention}
\end{figure}

The three angular differences used by Woldesenbet and Williams are not
interchangeable.  They label their images so that photons arrive first
from image \#1 and last from image \#4, as shown in Figure
\ref{fig:labeling_convention}.  

They define angular differences as follows:
\begin{eqnarray}\label{eqn:angle_diff_conv}
    \theta_{12} &= \pm(\theta_2 - \theta_1), \nonumber \\
    \theta_{23} &= \mp(\theta_3 - \theta_2), \nonumber \\
    \theta_{34} &= \pm(\theta_4 - \theta_3),
\end{eqnarray}
where $\pm$ is $+$ when the source position, $\phi_s$, is in
$2^\textrm{\scriptsize nd}$ or $4^\textrm{\scriptsize th}$ quadrant
and $-$ when $\phi_s$ in $1^\textrm{\scriptsize st}$ or
$3^\textrm{\scriptsize rd}$ quadrant.  Note that $\theta_{23}$,
the angular difference between the two closest images, is
defined oppositely from the other two.  Its modulus is always
less than $90^\circ$.  The directions in which the differences are
measured are shown in Figure
\ref{fig:labeling_convention}.

\begin{figure*}
\gridline{\fig{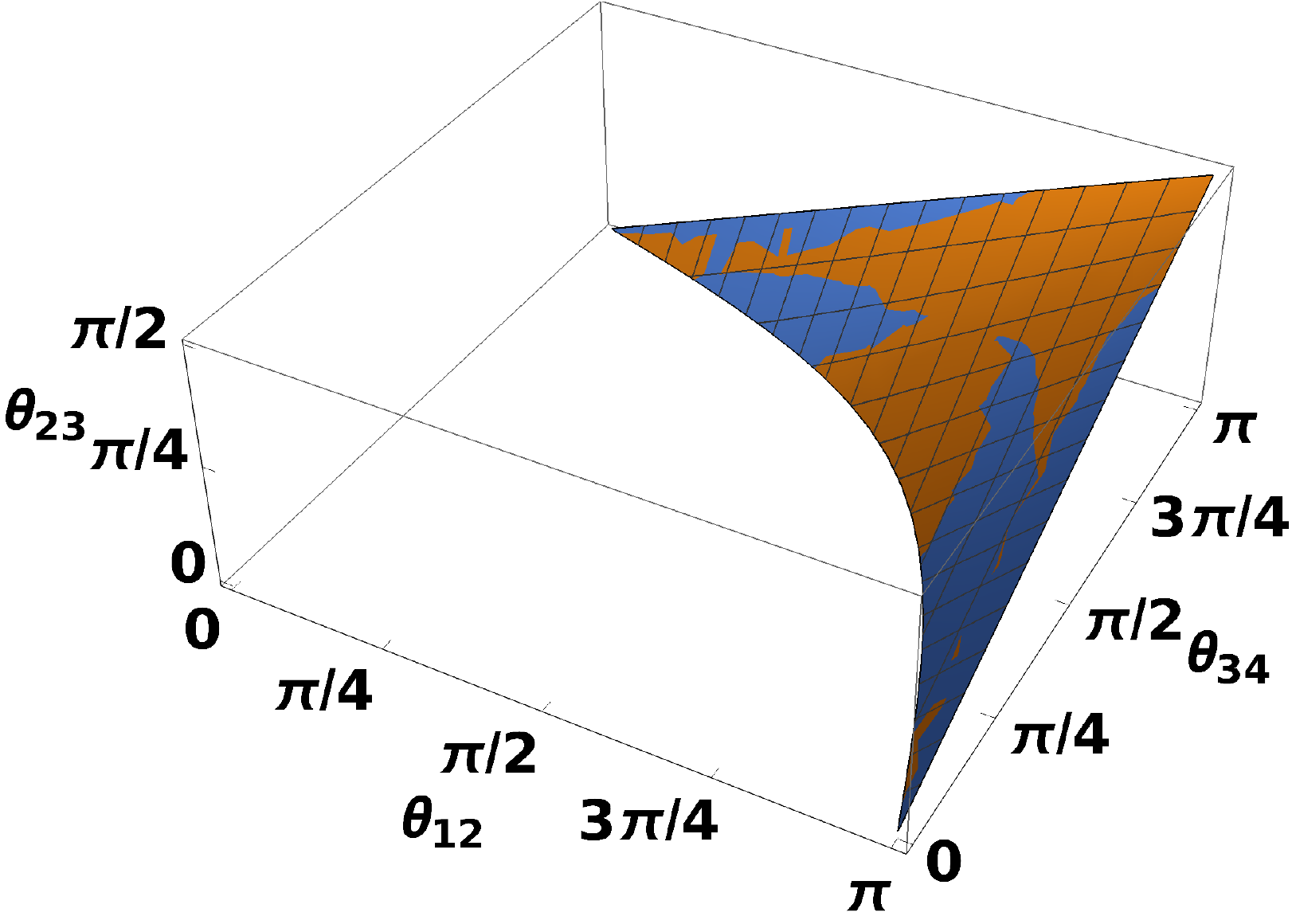}{0.49\textwidth}{(a)}
          \fig{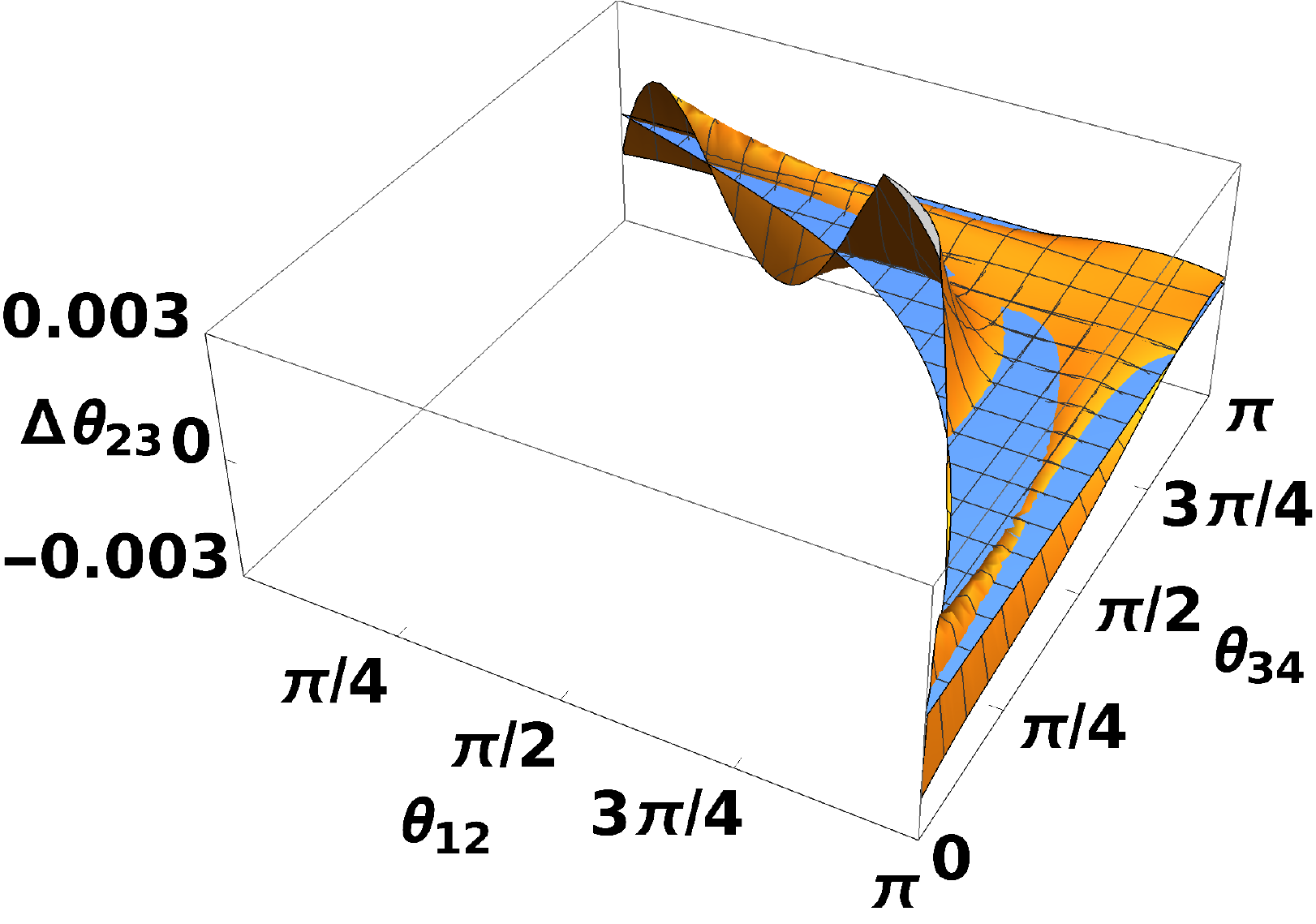}{0.49\textwidth}{(b)}}
\caption{ The two plots compare the Fundamental Surface of Quads (FSQ)
  (orange) and the surface corresponding to the configuration
  invariant (blue).  Panel (a) shows the FSQ and the surface of
  configuration invariant superimposed on each other, restricted to
  $\theta_{23} > 0$.  Panel (b) shows the difference between the FSQ
  and the configuration invariant. The blue is the zero plane and
  orange denotes the difference function. Note the different scales on
  the vertical axes. The planar axes are the same for both panels,
  while the vertical axes are $\theta_{23}$ and $\Delta
  \theta_{23}$, respectively.
\label{fig:KK_WW_plot}}
\end{figure*}
\subsection{The Fundamental Surface of Quads}
\cite{WilliamsWoldesenbet} discovered a quasi-two-dimensionality of
the space of quadruple lenses.
They used the closed form solutions for the  four image
positions, $\theta_i$, of the 
SIQP (which they call SISell) and after taking differences,
plotted them on the 
basis $(\theta_{23}, \theta_{12}, \theta_{34})$. They found a ``nearly
invariant two-dimensional surface" -- the Fundamental
Surface of Quads --  to which they fitted a polynomial
\begin{eqnarray}\label{FSQequation}
        \t_{23} &=&  -  5.792 + 1.783\: \t_{12} + 1.784\: \t_{34} \nonumber \\
        &+&0.1648\: \t_{12}^2 + 0.1643\: \t_{34}^2 -0.7275 \: \t_{12} \t_{34} \nonumber \\
        & - & 0.04591\:\t_{12}^3 -0.04579\:\t_{34}^3 \nonumber \\ 
        &+& 0.0549\: \t_{12}^2\t_{34} + 0.05493\: \t_{12}\t_{34}^2 \nonumber \\
        & - & 0.0001486\: \t_{12}^4 -0.0001593\: \t_{34}^4 \nonumber \\
        &+&0.01487 \: \t_{12}^3\t_{34} + 0.01487 \: \t_{12} \t_{34}^3 \nonumber \\
        &-& 0.03429 \: \t_{34}^2\t_{12}^2.
 \end{eqnarray}
This shows that the image configurations inhabit a space that is, to
good approximation, two dimensional.  Their invariant surface, expressed
as a power series of differences of angles, and the Kassiola and
Kovner configuration invariant, expressed in terms of directly
measured angles, are not obviously related.  But if 
there were two distinct invariants, the space of models
would be one dimensional. In the next subsection, we show that the surface
found by inverting the configuration invariant closely resembles the FSQ.

\subsection{Angular differences from the configuration invariant}

As image positions in the FSQ are expressed in terms of differences of angles, we should convert the configuration invariant accordingly to compare the two equations directly. 
Adopting the conventions of Equation (\ref{eqn:angle_diff_conv}), the configuration invariant, Equation (\ref{eqn:KKinvarianteq}), can be rewritten as
\begin{eqnarray}\label{modifiedKKeq}
    \cos\left(\theta_{23}-\frac{\theta_{12} + \theta_{34}}{2} \right) + \cos\left(\frac{-\theta_{12} -\theta_{34}}{2} \right) \nonumber  \\
     + \cos\left(\frac{\theta_{12} - \theta_{34}}{2}\right) = 0,\quad
\end{eqnarray}

\begin{equation}
        \cos\left(\frac{\theta_{12} + \theta_{34}}{2}-\theta_{23} \right) + 2\cos\left(\frac{\theta_{12}}{2}\right)\cos\left(\frac{\theta_{34}}{2}\right) = 0. 
\end{equation}

Solving for $\theta_{23}$,  we obtain the following equation
\begin{equation}\label{KK_theta_23_equation}
        \theta_{23} = \frac{\theta_{12} +\theta_{34}}{2} - \arccos\left(-2 \cos\frac{\theta_{12}}{2} \cos\frac{\theta_{34}}{2}\right).
\end{equation}

In Figure
\ref{fig:KK_WW_plot}(a)
we compare the polynomial Equation
(\ref{FSQequation}) for $\theta_{23}$, the angle between the closest
pair of images, with Equation (\ref{KK_theta_23_equation}), the
expression for the same angular difference derived from the
configuration invariant, over the range of $\theta_{12}$ and
$\theta_{34}$

Figure
\ref{fig:KK_WW_plot}(a)
shows how small the difference is
between the two results.  It would seem that \cite{WilliamsWoldesenbet}
re-discovered the configuration invariant, casting it in an
observer-friendly form that eliminated one dimension from the space of
observables and in the process obscured the connection.

\begin{figure}[ht]
    \centering
    \includegraphics[width=0.48\textwidth]{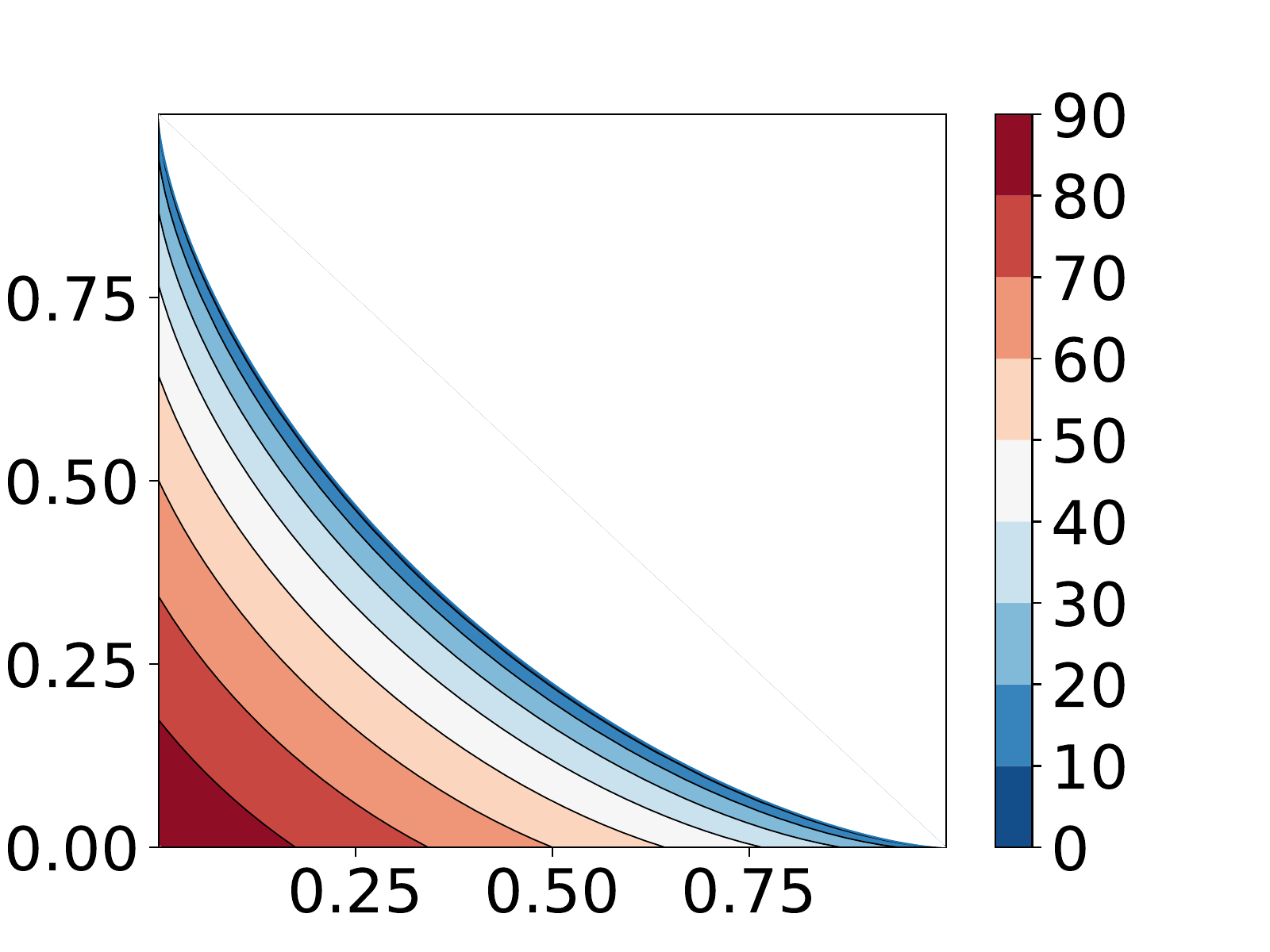}
    \caption{Contours of angle difference between the closest $2$ images over a quadrant of astroid caustic.}
    \label{plt:theta_differences_parametrization}
\end{figure}
\subsection{The FSQ and diamond caustic}
The two closest images merge when the source crosses the diamond
caustic, where $\theta_{23}=0$.  Figure
\ref{plt:theta_differences_parametrization} shows how $\theta_{23}$
approaches zero as the source approaches the diamond caustic.

The FSQ in Figure
\ref{fig:KK_WW_plot}(a)
intersects the $\theta_{23}=0$ plane in an
arc that looks very much like the arc in the lower left quadrant
of Figure
\ref{fig:tangent_hyperbola_astroid}

But the midpoint of the FSQ arc is at
$(\theta_{12},\theta_{34}) = (2\pi/3, 2\pi/3)$, putting it 33.3\% of
the way from the corner at $(\pi, \pi)$ to the corner at $(0,0)$.
By comparison, the midpoint of a unit astroid is
at $x^{2/3} = y^{2/3} = 1/2$, each of which is 35.3\% of the semi-axis of
the astroid.  Hence the shapes are subtly different.

\section{Semi-astroidal coordinates}\label{sec:semi-astroidal coordinates}
For both the SIEP and the SIQP we have shown that the ACLE has four
real solutions if complex quantity, $W$, defined, respectively,
in equations (\ref{eqn:WforSIEP}) and (\ref{eqn:WforSIQP})
is restricted to lie inside
a corresponding unit astroid.  For the SIEP the complex quantity is
the center of Witt's hyperbola.  For the SIQP, it is the source
position scaled by
half of the diameter of the  diamond caustic.
While the diamond caustic reduces to a point under the limiting case
of a vanishingly elliptical lens, the potential still forms a stable
image configuration if the position of the source relative to the diamond
caustic remains the same.  This suggests adopting specialized two
dimensional coordinate system with one coordinate constant on
concentric astroids.

\subsection{``Causticity" and position angle}\label{subsec:semi-astroidal coordinates}

A newly defined quantity should be given both a symbol and a name,
with the latter somehow evoking what is being quantified.  We name our
coordinates making specific reference to the case of a diamond
caustic, but note that they apply equally well to describe the
position of the Witt-Wynne hyperbola within {\it its} unit astroid.

We define a ``causticity," denoted by $\zeta$, that gives the relative
displacement of the source towards the astroid.  It is the ratio of
the size of the concentric astroid on which the source lies to the
size of the diamond caustic inside which four images will be produced.
If $W$ is given by the position of the source, as in equation (9),
then in a coordinate system aligned with the astroid, we define
\begin{equation}\label{eqn:zetadef}
\zeta \equiv \left[\left(\frac{x_s}{4 \epsilon b}\right)^{2/3} + \left(\frac{y_s}{4 \epsilon b}\right)^{2/3}\right]^{3/2}.
\end{equation}
Loci of constant causticity lie on similar concentric astroids.  For
the angular position of the source, we use $\phi_s$, which 
represents the angle that the source makes with respect to the
symmetry axis.  We call these the ``semi-astroidal" coordinates.


\begin{figure}[ht]
    \centering \includegraphics[width=0.5\textwidth]{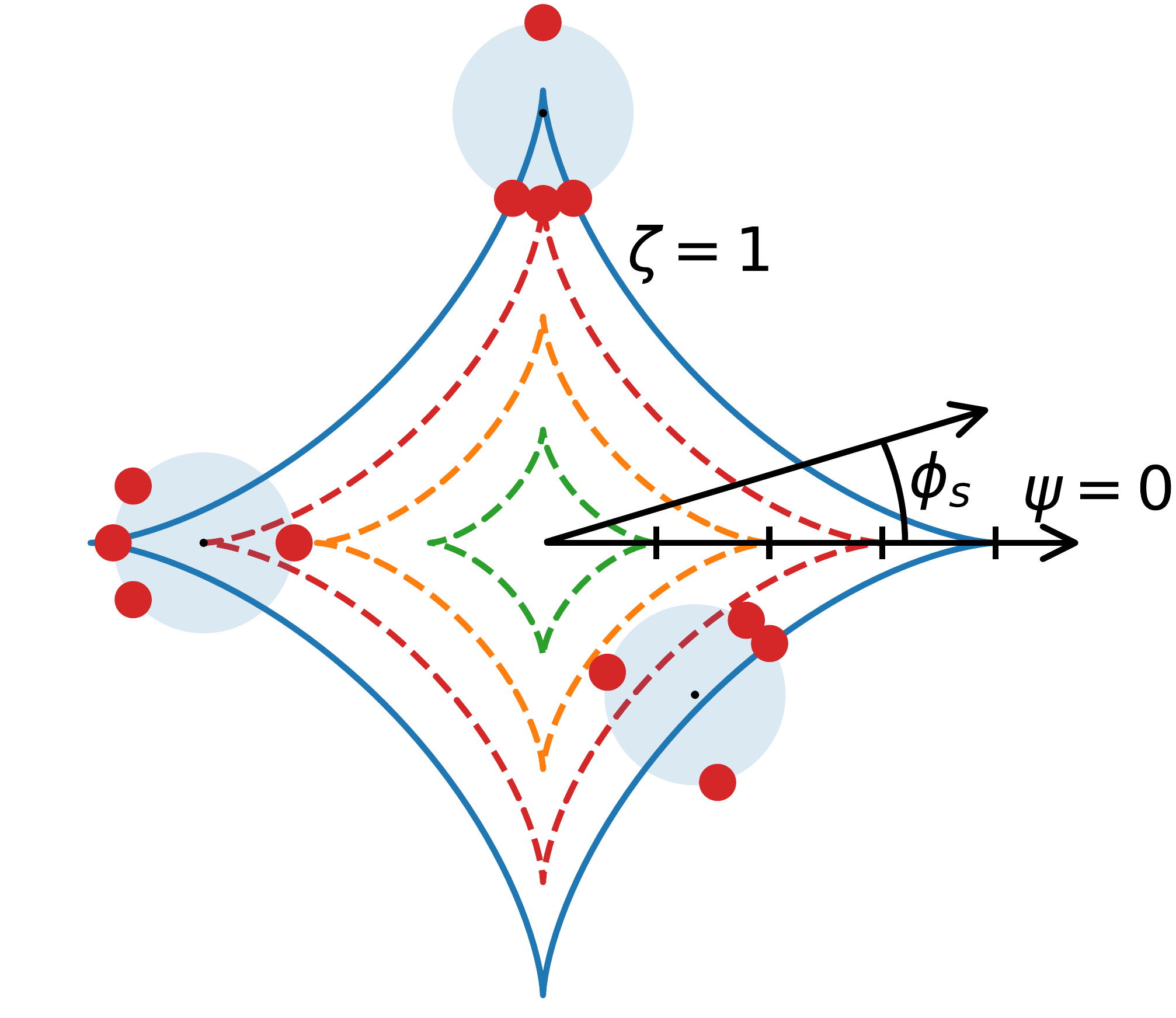} \caption{The
    red dots are quadruple configurations of lensed quasars arising
    from different source positions within a diamond caustic -- the
    blue astroid for which $\zeta = 1$.  Two of the pale blue circles
    are centered at $\zeta = 0.95$, with $\phi_s = \frac{\pi}{2}$ and
    $-\frac{\pi}{4}$, respectively.  The leftmost configuration has
    $(\zeta, \phi_s) = (0.75, \pi)$. The dashed curves are astroids
    of constant causticity, with $\zeta = 0.75$ (red), $\zeta =
    0.5$ (orange) and $\zeta = 0.25$ (green).}  \label{fig:astroidal_plot}
\end{figure}

Figure
\ref{fig:astroidal_plot}
shows several configurations corresponding
to different semi-astroidal coordinates and how they change as we
shift the position of the source within the caustic. \cite{Tuan_2020}
also separated the astroid caustic into areas corresponding to
different quads according to the \cite{Saha_2003} classification.

Though we have used the source position relative to a diamond caustic
to define our coordinates, there are corresponding definitions for the
center of the Witt's hyperbola with respect to the Witt-Wynne diamond.

\section{DE-SCRONCHED CONFIGURATIONS OF THE SIS+XS FROM THE ACLE} 
\label{sec: ACP from shear potential}


Until this point, our discussion has focussed on asymptotically
circular image configurations.  In our treatment of the
not-at-all-circular SIQP, we considered only its angular
configurations, which give rise to the ACLE despite its
non-circularity.

In this section we show that asymptotically circular configurations
can be used to produce the quadruple image configurations of a
different non-circular potential, the singular isothermal potential
with non-vanishing external shear (SIS+XS).

\begin{figure}[!b]
    \centering
    \includegraphics[width=8.5cm]{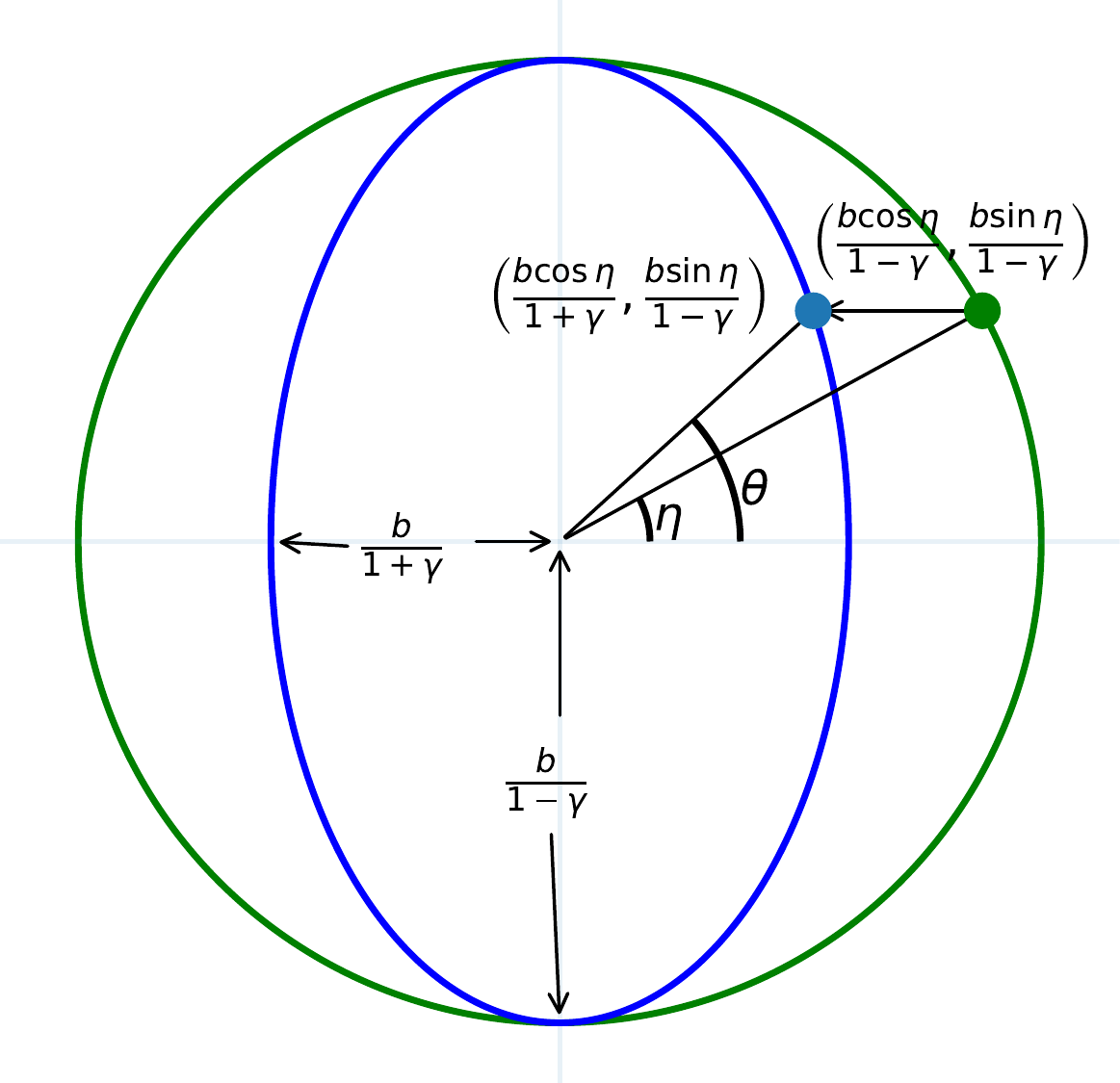}
    \caption{The ``scronched'' Wynne ellipse for an SIS+XS lens in blue
    and its  auxiliary circle in green.  The blue dot is an image formed by
     the SIS+XS at angle  $\theta$; the green dot shows its
     projection onto the auxiliary circle at
     $\eta$,  the ``eccentric angle''.
     }
     \label{fig:ecc_anom_plot}
\end{figure}

For the sake of concise description, we make frequent use of the word
``scronch," used by \cite{ellenberg2021shape} in his popular book
{\it Shape} to describe stretching in one direction and squeezing in the
orthogonal direction.  We show here that a quadruple image
configuration for an SIS+XS lens can be ``de-scronched" to produce
a corresponding asymptotically circular configuration.  

The argument proceeds as follows. \citet{Luhtaru_2021} have shown that
four images for an SIS+XS lens lie at the intersections of a Witt
hyperbola and a Wynne ellipse.  The positions of those four images,
projected onto the auxiliary circle circumscribing the Wynne ellipse
have ``eccentric angles'' $\eta$ that are shown in Section
\ref{subsec: angular equation identical for SIS+XS and ACP}
to be solutions of the
asymptotically circular lens equation.

The diamond caustic for an SIS+XS lens is a ``scronched" astroid.  For
the purpose of tying the positions of sources within these scronched
astroids to the positions of sources in the true astroids of their
associated asymptotically circular lenses, we define ``astroidal
coordinates", causticity and ``astroidal angle'' for a scronched
astroid in Section
\ref{subsec:astroidal coordinates}.
All SIS+XS configurations with the same causticity and astroidal angle
de-scronch to the same solution  of the ACLE. Given the shear of an SIS+XS
lens and the position of the source relative to its scronched astroid, one can
solve the lens equation by finding the image configuration for the
corresponding asymptotically circular lens and then scronching the
configuration.

\subsection{De-scronching configurations of the SIS+XS}
\label{subsec: angular equation identical for SIS+XS and ACP}

The potential for the singular isothermal sphere with finite external
shear centered on the lens\footnote{The sign convention for shear is
  chosen such that the symmetry axes coincide with those of the SIQP
  for $\gamma>0$. This choice produces equipotentials that are
  elongated in the $x$-direction.} (SIS+XS)
is given by,
\begin{equation}\label{eqn:SIS+XS potential}
    \Phi(x,y) = b\sqrt{x^2 +  y^2} - \frac{\gamma}{2}(x^2 - y^2) \quad,
\end{equation}
\noindent
with the lensing galaxy at
the origin of the coordinate system.  This differs from the case
considered by \citet{Luhtaru_2021} who, following \citet{Witt_1996},
expand the shear around the source position.  In Appendix
\ref{app:SIS+XS conics appendix}
we
derive expressions for Witt's hyperbola and Wynne's ellipse
appropriate to our lens-centered coordinate system.

Witt's hyperbola is given by
\begin{equation}\label{eqn:Witt's hyperbola}
    \frac{(1-\gamma)y-y_s}{(1+\gamma)x - x_s} = \frac{y}{x} \quad .
\end{equation}
\noindent
Wynne's ellipse is given by
\begin{equation}\label{eqn:Wynne's ellipse}
\left[{x - x_s/(1 + \gamma) \over b/(1 + \gamma)}\right]^2 +
\left[{y - y_s/(1 - \gamma) \over b/(1 - \gamma)}\right]^2 = 1
\quad . 
\end{equation}
\noindent
Note that it is centered {\it neither} on the source nor
on the lensing galaxy but at
\begin{equation}\label{eqn:ellcen}
   (x_e, y_e) =    \left({x_s \over[1 + \gamma]}, 
                         {y_s \over[1 - \gamma]}\right) \quad .
\end{equation}    
Taking the shear, $\gamma$ to be positive, Wynne's ellipse has
semi-major axis
 $b/(1 - \gamma)$
along the $y$-direction,
perpendicular to the elongation of the potential, semi-minor axis
 $b/(1 + \gamma)$
along the $x$-direction.  and axis ratio $(1-\gamma)/
  (1+\gamma)$.

Figure
\ref{fig:ecc_anom_plot}
shows the Wynne ellipse for an SIS+XS
configuration along with one of its images.  It also shows the
auxiliary circle with radius $b/(1 - \gamma)$ circumscribing that
ellipse.  Suppose $(x,y)$ is a solution of the lens for our SIS+XS
system, projecting to eccentric angle $\eta$ on the auxiliary circle.
Then
\begin{equation}\label{eqn:ptBUTxeye}
  (x - x_e, y - y_e )=\left(\frac{b \cos{\eta}}{1+\gamma}, 
  \frac{b \sin{\eta}}{1-\gamma}\right) \quad .
\end{equation}
Rewriting this is terms of the source position
using Equation (\ref{eqn:ellcen})
and substituting into the
numerator and denominator of 
the left hand side of 
Witt's hyperbola,
Equation (\ref{eqn:Witt's hyperbola}), gives
\begin{eqnarray}\label{lhs}
(1 - \gamma)y - y_s = b \sin\eta \quad {\rm and} \quad \nonumber \\
~~(1 + \gamma)x - x_s = b \cos{\eta}\quad ~~.~ \quad 
\end{eqnarray}
Subtituting 
into the numerator and denominator of its right hand side gives
\begin{eqnarray}\label{rhs}
y = (y_s + b\sin\eta)/(1 - \gamma) \quad {\rm and} \quad \nonumber \\
x = (x_s + b\cos\eta)/(1 + \gamma)\quad ~~.~~ \quad 
\end{eqnarray}
Combining these,
the image position on Wynne's ellipse,
$(x,y)$, 
can be eliminated from the equation
for Witt's hyperbola 
in favor of
its projected eccentric angle $\eta$.
It becomes,
\begin{equation}\label{hyperbola_ellipse_combined_equation}
\frac{\sin{\eta}}{\cos{\eta}} = \left(\frac{1+\gamma}{1-\gamma}\right)\left(\frac{y_s + b \sin{\eta}}{x_s +  b \cos{\eta}}\right) . \\
\end{equation}

After straightforward algebra detailed in Appendix
\ref{app:SIQP_SISXS_equation_relation},
this becomes the asymptotically circular lens equation,
\begin{equation}\label{eqn: ALCEXS}
    e^{4i\eta} - 2W e^{3i\eta}  + 2\bar{W}e^{i\eta} - 1 = 0 \quad ,
\end{equation}
with
\begin{equation}
  W = \left[\frac{(1-\gamma)x_s - i(1+\gamma)y_s}{2b\gamma}\right].
\end{equation}
The four solutions for $\eta$ give the eccentric angles on the
auxiliary ellipse associated with the four lens images on Wynne's
ellipse.  The unit astroid bounding $W$ has cusps on the horizontal
and vertical axes, respectively, at
\begin{equation}
x_s = \pm \left(\frac{2b\gamma}{1-\gamma}\right)~;~
y_s = \pm \left(\frac{2b\gamma}{1+\gamma}\right)\quad  , 
\end{equation}
which would indicate that the diamond caustic is a scronched
astroid elongated in the $x$-direction, perpendicular to Wynne's
ellipse, and given by
\begin{equation}\label{eqn:scronchedCAUSTIC}
    \left(
    \frac{(1-\gamma)x}{2\gamma b}\right)^{2/3} +
    \left(
    \frac{(1+\gamma)y}{2\gamma b}\right)^{2/3} = 1.
\end{equation}

We can alternatively define $W$ in terms of the position
of the center of Witt's hyperbola {\it relative} to the center
of Wynne's ellipse.  Starting with Equation (\ref{eqn:Witt's hyperbola})
we find 
\begin{equation}
    (x_h, y_h) = \left(\frac{x_s}{2\gamma}, -\frac{y_s}{2\gamma}\right)
    \quad . 
\end{equation}
Subtracting Equation (\ref{eqn:ellcen})
\begin{eqnarray}
x_h - x_e =
~~\left(\frac{1-\gamma}{1+\gamma}\right)\frac{x_s}{2\gamma} \quad \rm{and}\quad
\nonumber \\
  y_h - y_e =
-\left(\frac{1+\gamma}{1-\gamma}\right)\frac{y_s}{2\gamma} ~~ \quad ~.~ \quad
\end{eqnarray}
giving
\begin{equation}\label{eqn:WforSISXS}
  W = \left[
      \frac{x_h - x_e}{b/(1+\gamma)}
  + i \frac{y_h - y_e}{b/(1-\gamma)}\right].
\end{equation}
The unit astroid bounding $W$ has cusps on the horizontal
and vertical axes, respectively, 
\begin{equation}
x_h - x_e = \pm \frac{b}{(1+\gamma)} ~;~ y_h - y_e = \pm \frac{b}{(1-\gamma)} , 
\end{equation}
which
indicates that the Witt-Wynne diamond 
is a scronched astroid elongated in the $y$-direction,
\begin{equation}\label{eqn:scronchedWW}
    \left(
    \frac{x}{b/(1+\gamma)}\right)^{2/3} +
    \left(
    \frac{y}{b/(1-\gamma)}\right)^{2/3} = 1,
\end{equation}
which is 
{\it perpendicular} to the diamond caustic,
Equation (\ref{eqn:scronchedCAUSTIC}) and
{\it parallel} to Wynne's ellipse.
\begin{figure*}[t]
  \includegraphics[scale=0.6]{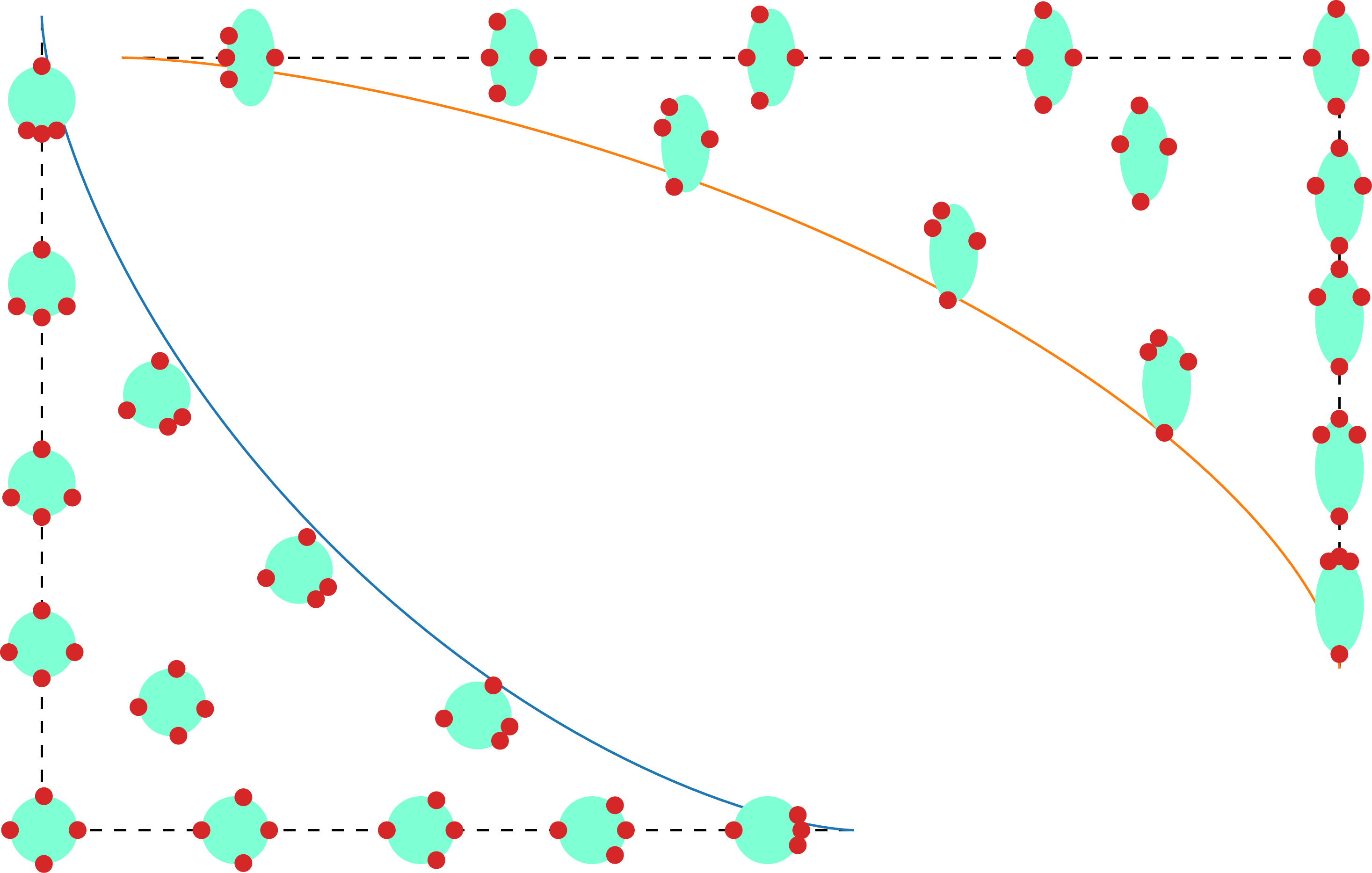}
  \caption{Image configurations for SIS+XS lens systems with the
    source at various positions in one quadrant of the diamond
    caustic.  The cyan Wynne circles and Wynne ellipses are centered
    at the source position.  The lower left shows the limiting case of
    shear $\gamma \rightarrow 0$, while the upper right shows
    $\gamma= 1/3$, and has been rotated $180^\circ$.  The
    $\gamma \rightarrow 0$
    diamond caustic is scronched {\it parallel} to the elongation of
    the potential to produce the $\gamma = 1/3$ diamond caustic.  The
    $\gamma = 1/3$ Wynne ellipses are scronched {\it perpendicular} to
    the elongation of the potential and the diamond caustic.  Apart
    from the $180^\circ$ rotation, the causticities $\zeta$ and astroidal
    angles $\alpha$ are the same for both sets of configurations.}
\label{fig:sheared_astroidal_plot}
\end{figure*}

\subsection{Astroidal coordinates}
\label{subsec:astroidal coordinates}
\citet{Finch_2002}\footnote{They use the opposite convention for
shear. Their equations have been modified so that the potential
is elongated in horizontal direction.}
represent the locus of the SIS+XS diamond caustic parametrically,
in terms of $\theta_c$, the polar coordinate
in the image plane of the point on the critical curve at which two
images converge when a source crosses the associated point on
the caustic.  The locus of the caustic traced by $\theta_c$ is given by 
\begin{equation}
   x_a = \frac{2b\gamma}{1-\gamma} \cos^3 \theta_c ~;~ y_a = -\frac{2b\gamma}{1+\gamma} \sin^3 \theta_c \quad .
\end{equation}
Note that
\begin{equation}\label{tanPHI}
\tan \phi_s = \frac{y_a}{x_a} = -\frac{1-\gamma}{1+\gamma}\tan^3 \theta_c\quad .
\quad
\end{equation}

As was evident in Equation (\ref{eqn:scronchedCAUSTIC}), the diamond
caustic is a scronched astroid, stretched by a factor of 
$\frac{1}{1 - \gamma}$ in the horizontal direction and squeezed by a
factor of $\frac{1}{1+\gamma}$ in the vertical direction.  As we
de-scronch the diamond caustic, maintaining the relative position of
the source, $\phi_s$, varies as ${(1-\gamma)\over(1+\gamma)}$.  Thus
$\theta_c$ remains unchanged.

The principal result of this section can be succinctly expressed
by introducing coordinates that are invariant under scronching.
We define an ``astroidal angle'', $\alpha$, such that 
\begin{equation}
\tan^3 \alpha \equiv \left(\frac{1 + \gamma}{1 - \gamma}\right) \tan  \phi_s \quad .
\end {equation}
so that $\alpha = -\eta_c$.

We also extend our original definition of causticity, Equation
(\ref{eqn:zetadef}) to include non-vanishing shear, 
\begin{equation}\label{eqn:newzetadef}
\zeta \equiv \left[
\left(\frac{(1 - \gamma) x_s}{2 b \gamma }\right)^{2/3} +
\left(\frac{(1 + \gamma) y_s}{2 b \gamma }\right)^{2/3}\right]^{3/2} , 
\end{equation}
noting that $\zeta$ is unchanged as one de-scronches the diamond
caustic.

We conclude that the image configuration for an SIS+XS lens is a
scronched version of the image configuration for the corresponding
de-scronched circular lens with the same causticity $\zeta$ and
astroidal angle $\alpha$.

\subsection{The scronching of the asymptotically circular
configurations and their diamond caustic}

In Figure
\ref{fig:sheared_astroidal_plot}
we show thirteen quadruple
image configurations of the asymptotically circular lens and the
corresponding configurations of an SIS+XS lens with identical
astroidal coordinates.  Each configuration is centered at the position
of the source within the diamond caustic.  The asymptotically circular
astroid is scronched by the same factors as the Wynne circle. but with
the axes switched.

\section{ACLE SOLUTIONS FOR DE-SCRONCHED CONFIGURATIONS OF THE SIEP} 
As with the SIS+XS, the image positions for
the singular isothermal elliptical potential (SIEP) with non-vanishing
ellipticity can {\it also} be found by de-scronching Wynne's ellipse
to a circle, solving the asymptotically circular lens equation,
and then scronching the configuration.

For the sake of comparision with the SIS+XS, we use semi-ellipticity
$\hat \epsilon$ \citep{Luhtaru_2021} rather than axis ratio $q$ where
\begin{equation}
 \hat\epsilon \equiv \frac{1 - q}{1 + q} \quad .
\end{equation}
The SIEP is then
\begin{equation}
\Phi(x,y)= \sqrt{[(1-\hat \epsilon)x]^2 + [(1+\hat\epsilon)y]^2} \quad .
\end{equation}

The diamond caustic of the SIEP is {\it not} a scronched astroid, but
 the positions of its four images can nonetheless be found at the
 intersection of a Wynne ellipse and a Witt hyperbola \citep{Luhtaru_2021}.  We show here that the Witt-Wynne diamond {\it is} a
 scronched astroid, as was the case for the SIS+XS.

Wynne's ellipse is elongated perpendicular to the long axis
of the potential and is given by
\begin{equation}\label{eqn:wynneSIEP}
\left[{x-x_s \over b/(1 + \hat\epsilon)}\right]^2 +
\left[{y-y_s \over b/(1 - \hat\epsilon)}\right]^2 = 1
\quad . 
\end{equation}
This differs trivially from Wynne's ellipse the SIS+XS, Equation (\ref{eqn:Wynne's ellipse})
in the replacement of the shear,  $\gamma$ with the
semi-ellipticity, $\hat \epsilon$, but 
non-trivially in that, in contrast to the SIS+XS, the ellipse
for the SIEP {\it is} centered on the source position.

Witt's hyperbola is given by 
\begin{eqnarray}
  [(y-y_s)-(y_h - y_s)]&[(x-x_s)(x_h - x_s)] \nonumber \\
  & = a^2/2 \quad , \label{eqn:righthyperbola}
\end{eqnarray}
where $a$ is the semi-major axis of the hyperbola.  It can
be evaluated by substituting the center of 
Wynne's ellipse, $(x_s, y_s)$, which must, by construction,
lie on Witt's hyperbola, for $x$ and $y$ in Equation 
(\ref{eqn:righthyperbola}),
giving
\begin{equation}
a^2/2 = (y_h - y_s)(x_h - x_s) \quad .
\end{equation}
Witt's hyperbola then simplifies to
\begin{eqnarray}
(y-y_s)(x-x_s)-(y-y_s)(x_h-x_s) \quad \quad \quad  \nonumber \\
-(x - x_s)(y_h-y_s) = 0.\quad
\end{eqnarray}

Though it was introduced for the SIS+XS, Figure
\ref{fig:ecc_anom_plot}
can
equally well be applied to the SIEP.
If $(x,y)$ is a solution of the lens equation for a SIEP
system, projecting to eccentric angle $\eta$ on the auxiliary circle
gives
\begin{equation}
  (x - x_s, y - y_s )=\left(\frac{b \cos{\eta}}{1+\hat\epsilon}, 
  \frac{b \sin{\eta}}{1-\hat\epsilon}\right) \quad .
\end{equation}

Substituting this general point on the Wynne's ellipse into the equation
for Witt's hyperbola, we get 
\begin{eqnarray}
 \sin\eta \cos \eta & = & - (1 + \hat \epsilon)\sin \eta (x_h - x_s)/b \nonumber \\
                    &  & - (1 - \hat \epsilon)\cos \eta (y_h - y_s)/b~.
\end{eqnarray}                    

Dividing by $\sin\eta \cos\eta$ and letting
\begin{eqnarray}
p & = & (1 + \hat \epsilon) (x_h - x_s)/b~~~{\rm and}\\
q & = & (1 - \hat \epsilon) (y_h - y_s)/b~~~~~.
\end{eqnarray}
we have
\begin{equation}
1 = p \sec \eta + q \csc \eta \quad ,
\end{equation}
which is exactly the deceptively simple form of the ACLE, Equation
(\ref{eqn:deceptive}).  In the defining form of the ACLE,
Equation (\ref{eqn:Main_ACP_equation}), the quantity W
becomes
\begin{equation}
  W = \left[
      \frac{x_h - x_s}{b/(1+\hat\epsilon)}
  + i \frac{y_h - y_s}{b/(1-\hat\epsilon)}\right].
\end{equation}
The corresponding value for the SIS+XS, Equation (\ref{eqn:WforSISXS}) differs only
in that it explicitly depends on the center of Wynne's ellipse,
$(x_e,y_e)$ rather than implicitly  through the source position.

As with the SIS+XS, the Witt-Wynne diamond is a scronched astroid
elongated parallel to the Wynne ellipse and perpendicular to the
potential.  Unlike the case for the SIS+XS, the diamond caustic is not
a scronched astroid.

\section{Generalization to all circularly symmetric potentials}\label{circular_universality_section}
Until now we have analyzed the properties of circular configurations
as limiting cases of specific potentials. However, these properties
also hold more generally for all circularly symmetric
potentials.  Specifically, \cite{An_2005} showed that the caustic locus
depends only on ``the azimuthal behavior of perturbation of the
potential." He then showed that potentials with ellipticity, external
shear, and quadrupole moment can all be shown to have perturbations
of identical form to first order in the limit of vanishing
quadrupole.

This implies that the angular configurations (and the ratio of
magnifications) are identical for potentials with any of the three
abovementioned deviations from circularity, up to linear order in the
perturbations.

In this section we show that a generic circular potential with
vanishing external shear gives rise to a Witt hyperbola, with images
forming where that hyperbola crosses a 1D locus, the limiting
Wynne circle.

Our generic circular potential with shear can be written 
\begin{equation}
    \Phi(x, y) = f(x^2+y^2) - \frac{\gamma}{2}(x^2-y^2).
\end{equation}
Using the lens equation\citep{Bourassa_1975},
\begin{equation}
    \mathbf{r} -\mathbf{r}_s = \nabla{\Phi(\mathbf{r})},
\end{equation}
we get
\begin{eqnarray}
 x-x_s & = & 2xf'(x^2+y^2) -\gamma x\label{lens_equation_x_component1}\quad{\rm and}\\
 y-y_s & = & 2yf'(x^2+y^2) + \gamma y\label{lens_equation_y_component1} \quad .
\end{eqnarray} 

Rearranging and dividing Equation (\ref{lens_equation_y_component1}) by Equation (\ref{lens_equation_x_component1}), we obtain the equation of Witt's hyperbola, on which all four images should lie,
\begin{equation}
    \frac{(1-\gamma)y-y_s}{(1+\gamma)x - x_s} = \frac{y}{x}.
\end{equation}

This is exactly Witt's hyperbola for an SIS+XS potential. By rearranging and squaring Equations (\ref{lens_equation_x_component1}) and (\ref{lens_equation_y_component1}), we can obtain a one dimensional locus that reduces to
Wynne's ellipse for the case of SIS+XS,
\begin{eqnarray}
    ((1+\gamma)x - x_s)^2 + ((1-\gamma)y - y_s)^2 \nonumber\\= 4\gamma^2(x^2+y^2)(f'(x^2+y^2))^2\label{point_mass_elliptical_eq}
\end{eqnarray}

In the limiting case of vanishing shear,  ($\gamma \rightarrow 0, x_s \rightarrow 0, y_s \rightarrow 0$), and after taking the square root, 
Equation (\ref{point_mass_elliptical_eq}) becomes
\begin{equation}
 f'(x^2+y^2) = 1/2 \quad .
\end{equation}
Thus images form on a circle with squared radius $a^2 = x^2 + y^2$. 
\begin{figure}[ht]
    \centering
    \includegraphics[width=0.25\textwidth]{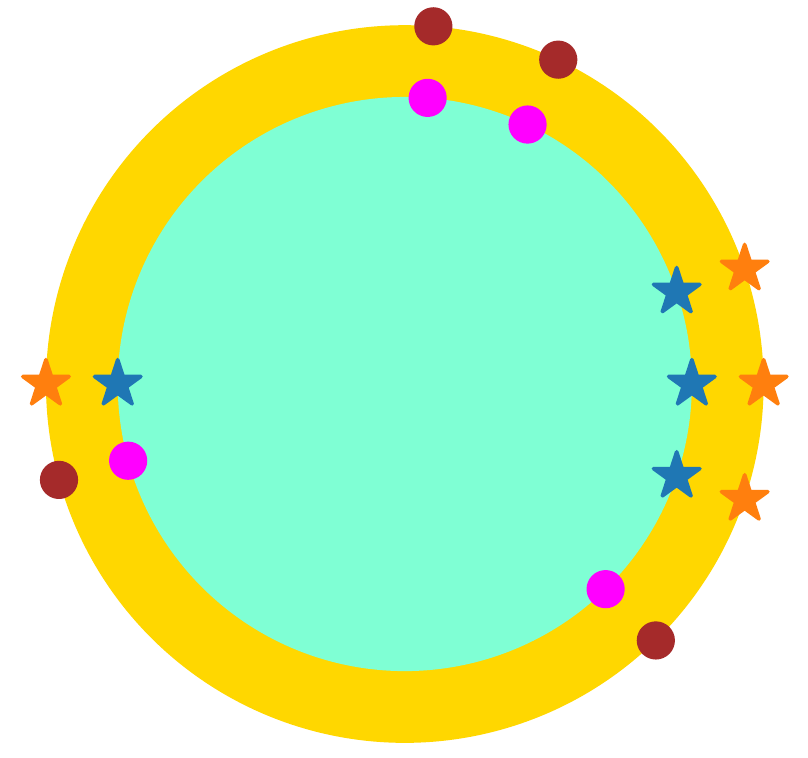}
    \caption
{Outer circle: images for two different source positions
for 1 point mass lens with asymptotically vanishing shear
and lens strength $b=1.25$.  Inner circle: images
formed by an SIS lens with vanishing shear ($\gamma\rightarrow 0$) and $b = 1$,
The source positions have the same 
causticity and astroidal angle for both circles, but 
the symmetry axes of the two configurations have been rotated to
avoid confusion.  Keeton's \texttt{lensmodel} \citep{Keeton_2010} has
been used to verify the configurations.}  \label{ptms_vs_SIQP_plot}
\end{figure}
As the configurations for an asymptotically circular lens are given by
the intersection of a circle and a hyperbola, the positions are
exactly the same as those we previously encountered for isothermal
potentials in the limiting circular case.

Figure
\ref{ptms_vs_SIQP_plot}
shows the predicted positions for a point mass lens and the SIS+XS potential in the limiting case.

\section{Flux ratios for the images in asymptotically circular configurations}\label{magnifications section}
As the area of the caustic decreases, the inverse magnification tends
to $0$, and hence, the brightnesses of all 4 images grow infinite.
It is nonetheless possible to calculate the ratios of the
brightnesses of the 4 images in the limiting case
of circularity.

In
Appendix
\ref{app:SISXS_magnification}
we derive the expressions for
magnifications of images formed by the SIQP and the SIS+XS potential
with non-vanishing quadrupole.  In asymptotically circular cases, with either $\gamma \rightarrow 0$
or  $\epsilon \to 0$ and $s \rightarrow 0$,
the magnification simplifies to
\begin{eqnarray}
\mu_{SIQP} &  =  &\frac{b}{s \cos(\t-\phi_s)-4\epsilon b \cos(2\t)}~{\rm and} \quad \quad \label{mag_SIQP}
\\
\mu_{SIS+XS} &  = & \frac{b}{s \cos(\t-\phi_s)-2\gamma   b \cos(2\t)}\quad , \label{mag_SISXS}
\end{eqnarray}
where $(s, \phi_s)$ are the polar coordinates of source, and $\t$ is the
polar angle of an image on the limiting circle.

We also show in Appendix
\ref{app:SISXS_magnification}
that if an SIQP
lens system and an SIS+XS lens system produce the same
asymptotically circular configuration, the quadrupole parameter
$\epsilon$, of the SIQP is equal to the shear parameter, $\gamma$ of the
SIS+XS.  The
magnifications for the SIS+XS system, Equation
(\ref{mag_SISXS}),
are then {\it twice} the magnifications
for the corresponding SIQP system, Equation
(\ref{mag_SIQP}).
The SIS+XS
has a diamond caustic that is {\it half} the area of the diamond
caustic for the SIQP (both of which are true astroids in the limit
of circularity).

\cite{An_2005} showed that for
small perturbations, the magnification ratios depend only on the
angular behavior of the potential, but that  the magnifications are scaled by
a different multiplicative factors for different spherically symmetric mass
distributions.\\
\begin{figure}[ht]
    \centering
    \includegraphics[width=0.5\textwidth]{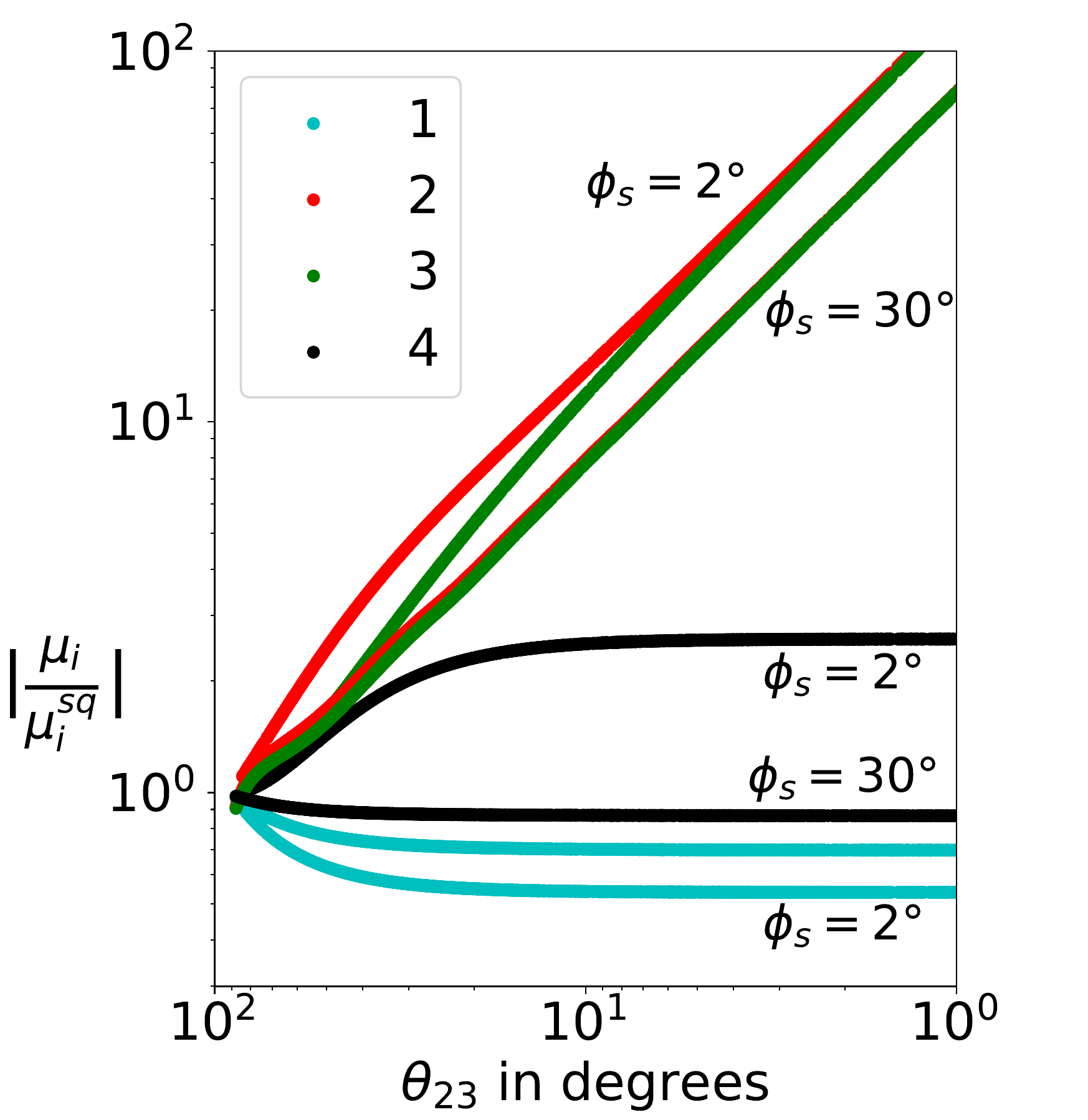}
    \caption{
Image magnifications for asymptotically circular image configurations,
relative to the image magnifications for a source at the center of
the diamond caustic.  Two cases are shown, one with the source
close to an axis of the diamond caustic $(\phi_s = 2^\circ)$,
and a second far from either axis $(\phi_s = 30^\circ)$.
The two closest images (those on the secondary branch of Witt's
hyperbola) increase in brightness inversely as their
separation.   By contrast, the relative magnifications
of the images at $\theta_1$ and $\theta_4$ on the {\it primary} branch of Witt's hyperbola 
vary only by factors of
order unity and quickly reach limiting values, even in the cusp-like case of
$\phi_s =  2^\circ$.  The 
convention is that of Section 4.1.
}
\label{fig:magnification_ratio}
\end{figure}
Figure
\ref{fig:magnification_ratio}
shows how magnification varies
for images formed by an asymptotically circular potential as $\t_{23}$ (the angle between the
closest images) varies. The magnifications of images $2$ and $3$
asymptote to a straight lines of unit slope in the inverted log-log
graph. This follows from the theorem that the magnification is
approximately inversely proportional to the angular separation between
the images \citep[p. 190]{schneider1992gravitational}.

\section{Solar System Occultation Flashes and the ACLE}
The wide applicability of asymptotically circular image configurations
extends even beyond gravitational lensing.

\cite{Nicholson_1995} noted the 
parallel between gravitational lensing and the ``central flash'' observed
in their observations  of Saturn's occultation of 28 Sgr on 3 July 1989.
The star never completely disappears behind the planet.  A refracted
image on Saturn's limb grows increasingly faint following ingress.
At the same time a second, even fainter, refracted image
brightens on the opposite limb.  Toward mid-occultation a bright new
pair of images forms causing the first of two flashes.  The images grow
fainter as they move apart, but one of them grows brighter again just
before it merges with the original image, which also grows brighter.

Their observations were sufficiently sensitive to show the two
bright images of the newly created pair but not the fainter images,
whose positions they calculate based on an assumed atmosphere.

Their Figure 4 shows theoretical calculations of a series of
configurations over the course of their observations.  We measured the
positions of the images directly from the figure and then de-scronched
them (taking Saturn's limb to be elliptical) to an auxiliary circle,
as in
Section 6.

The two larger eccentric angle differences, $\eta_{12}$ and $\eta_{34}$
were used to ``predict'' the smallest eccentric angle difference,
$\eta_{23}$ using Equation \ref{KK_theta_23_equation}.  The deviations
of the measured $\eta_{23}$ from the calculated $\eta_{23}$ have a
mean of $-0.455^\circ$ with a standard deviation of $1.04^\circ$.

By comparison, \cite{WilliamsWoldesenbet} found that only 12 out of 40
observed quads lie within $2^\circ$ of the Fundamental Surface of Quads.
This raises the possibility that the image configurations associated with
occultation flashes may be more like those expected for quadruply
lensed quasars than the quasar configurations themselves.
\vfill
\eject
\section{Discussion and Conclusions}

We have examined the circular image configurations of quadruply lensed
sources.  These occupy a two-dimensional space bounded by a suitably
defined unit astroid.  We define astroidal coordinates, ``causticity'', $\zeta$,
and ``astroidal angle'', $\alpha$, on this space. 

We considered a variety of distinct cases that give rise to such
configurations, first nearly-circular isothermal potentials with
vanishing ellipticity or shear and then nearly-circular
non-isothermal potentials with vanishing shear.  Across
these different cases, systems with identical astroidal
coordinates produce identical angular configurations.

We discussed the non-circular singular isothermal quadrupole lens, and
showed that it produces the same set of {\it angular} configurations
(with angles measured from the position of the lensing galaxy) even
though the images do not lie on a circle.

We extended our analysis to singular isothermal potentials of
non-vanishing ellipticity and circular isothermal potentials with
non-vanishing shear and showed that their image configurations are
``scronched'' versions of the asymptotically circular configurations --
stretched in one direction and squeezed in the other, as in
Figure
\ref{fig:sheared_astroidal_plot}.
We extended
our definitions of astroidal coordinates to incorporate these,
and found that systems with identical $\zeta$ and $\alpha$
have identical angular configurations when ``de-scronched''.

The Witt-Wynne \cite{} geometric solution for the SIEP helps to explain
the
frequent resurfacing of these configurations.  Witt found
that for non-vanishing elliptical potentials and for singular isothermal
potentials with external shear, the four images must lie on a hyperbola.
Wynne \cite{} found that the images for both cases must lie on an ellipse
if the potentials are isothermal.

Images formed at the intersection of a Witt hyperbola and a Wynne
ellipse of vanishing ellipticity (a circle) give an asymptotically
circular image configuration.  ``Scronching'' that circle to an ellipse
gives a configuration for a lens with non-vanishing quadrupole.

\begin{acknowledgments}
  C.\ F.\ gratefully acknowledges support from the MIT UROP office.
  The authors thank Professors L.\ Williams, P.\ Saha and C.\ Kochanek
  for their comments on an early version of the manuscript.
\end{acknowledgments}

\appendix

\section{The ACLE as an edge case of the Witt-Wynne construction}\label{app:ACPACE}

We derive the asymptotically circular lens equation (ACLE) from the
limiting case of the Witt-Wynne construction.  In this geometric
scheme, the four image positions lie at the intersections of a circle
and a hyperbola. To simplify the situation, we can translate and
rescale our coordinate system, such that the Wynne ellipse is a circle
of radius $b$ centered at the origin and the hyperbola passes through
the origin, as it should pass through the center of the circle.

Let the images be at polar angles $\theta_1, \theta_2, \theta_3,$ and
$\theta_4$. For simplicity, rotate the configuration such that
the major axis of the potential is at $\psi =
0$.  The general equation of a rectangular hyperbola
passing through the origin whose asymptotes are aligned with the
coordinate axes is
\begin{equation}
    (x-x_h)(y-y_h) = x_hy_h
\end{equation}

Substituting a general point from the circle, ${(b\cos\theta, b\sin \theta)}$,
we find
\begin{equation}\label{eqn:intermediate}
 b^2\sin\theta\cos\theta - b x_h \sin\theta - b y_h\cos\theta=0 \quad
\end{equation}
and hence
\begin{equation}\label{eqn:compact}
 1 = \frac{x_h}{b}\sec\theta + \frac{y_h}{b}\csc\theta \quad .
\end{equation}
This is the more compact of our two versions of the ACLE.

Defining $p \equiv \frac{x_h}{b}$ and $q \equiv \frac{y_h}{b}$, one
can show that $|p\sec \t + q\csc \t|$ attains a local minimum with
value $(p^{2/3}+q^{2/3})^{3/2}$.  Equation (\ref{eqn:compact}) will therefore have $4$
distinct solutions when
\begin{eqnarray}
    (p^{2/3}+q^{2/3})^{3/2} < 1 \quad {\rm or} \nonumber \\
    p^{2/3}+q^{2/3} < 1 ~~~, ~
\end{eqnarray}
and $2$ real solutions when the inequality is reversed.
Two of the solutions merge to a single solutions when equality holds.
Curves of constant  $p^{2/3}+q^{2/3}$ trace similar astroids in
the $(p,q)$ plane.

From equation (\ref{eqn:intermediate}) we obtain
\begin{equation}
    b^2\sin2\theta -2 b x_h \sin\theta - 2b y_h \cos\theta = 0,
\end{equation}

The four polar angles of the lens configuration,
$\theta_1, \theta_2, \theta_3,$ and $\theta_4$ are the solutions of
this equation. Replacing $\cos \alpha$ with $\frac{e^{i\alpha} + e^{-i\alpha}}{2}$ and $\sin \alpha$ with $\frac{e^{i\alpha} - e^{-i\alpha}}{2i}$, we obtain the equation:
\begin{eqnarray}\label{eqn:WS_configurational_equation}
    b^2 \left(\frac{e^{2i\t}-e^{-2i\t}}{2i}\right) - 2b x_h\left( \frac{e^{i\t}-e^{-i\t}}{2i}\right) - 2b y_h \left(\frac{e^{i\t}+e^{-i\t}}{2}\right) = 0\nonumber \\
    e^{4i\theta} - 2\left(\frac{x_h+iy_h}{b}\right) e^{3i\theta} + 2\left(\frac{x_h - iy_h}{b}\right)e^{i\theta} -1 = 0  \label{eqn:Wynne-Schechter-edge-case}
\end{eqnarray}
Note that if we let
\begin{eqnarray}
     W = \frac{x_h + iy_h}{b},
\end{eqnarray}
Equation (\ref{eqn:Wynne-Schechter-edge-case}) becomes
\begin{equation}
    e^{4i(\t-\psi)} - 2W e^{3i(\t-\psi)} + 2\overline{W}e^{i(\t-\psi)} - 1 = 0,
\end{equation}
where $\t$ is replaced with $\t-\psi$ for cases when $\psi \neq 0$.
This our second, more useful version of the ACLE.

Taking $\psi = 0$ for the sake of clarity, we can apply Vieta's formula to the coefficient of $e^{3i\t}$ term,
\begin{equation}\label{eqn:Vieta's formula sum of roots}
    \sum_{j=1}^4 e^{i\t_j} = 2W = 2\left(\frac{x_h + iy_h}{b}\right).
\end{equation}
So, the center of Witt's hyperbola can be expressed as
\begin{eqnarray}
    x_h = \frac{b}{2} \sum_{j=1}^4 \cos\t_j = 2 \sum_{j=1}^4 \frac{x_j}{4} = 2{x_\textrm{\scriptsize centroid}}\nonumber\\
    y_h = \frac{b}{2} \sum_{j=1}^4 \sin\t_j= 2 \sum_{j=1}^4 \frac{y_j}{4}= 2{y_\textrm{\scriptsize centroid}}
\end{eqnarray}

\section{Closed form solution of the ACLE}\label{app:closedform}
The  Asymptotically Circular Lens Equation (\ref{eqn:Main_ACP_equation}),
\begin{equation}
    e^{4i\theta} - 2W e^{3i\theta} +2\overline{W} e^{i\theta} - 1= 0 \quad,
\end{equation}
is quartic in the quantity $z=e^{i\theta}$.

Ferrari's oft-deprecated method can be used to find four closed form
solutions.  One juggles the terms in the equation (adding quantities
to both sides as needed), so that one has perfect squares on both
sides: a quadratic expression in $z$ squared on one side and
expression linear in $z$, also squared, on the other.  One takes the
square root and adds (and subtracts) the quadratic and linear
expressions, giving two new quadratics in $z$, which one then solves
giving four roots.

There is art in the juggling, creating the two perfect squares.  Almost
without exception, it is never possible using only the terms in the
original quartic.  But if one judiciously adds terms that include an
unknown algebraic quantity to both sides, one can solve for values of
that quantity that give two perfect squares.

We begin by isolating the $z^4$ and $z^3$ terms on one side of the
equation,
\begin{equation}
z^4 - 2W z^3 = -2 \overline{W} z + 1   \quad .
\end{equation}
Adding $W^2z^2$ to both sides gives us 
\begin{equation}\label{eqn:perfect}
 z^4 - 2Wz^3 + W^2z^2  = (z^2 - Wz)^2  =   W^2z^2 - 2 \overline{W} z + 1 \quad .
\end{equation}

The left hand side is a perfect square of a quadratic, but the
right hand side is not a perfect square of a linear function.
We add terms involving the algebraic quantity $u$, to both sides,
\begin{equation}\label{eqn:completedsquare}
 \left(z^2-Wz + \frac{u}{2}\right)^2 =
  (W^2+u)z^2-(2\overline{W}+ Wu)z+\left(1+\frac{u^2}{4}\right)\quad .
\end{equation}
The left hand side is again a perfect square, for all values
of $u$.  The right hand side {\it might} be a perfect square for some
{\it specific} values of $u$.

If such values exist, there is a constant $k$ such that
\begin{equation}\label{eqn:linearsquared}
  (z\sqrt{W^2 + u} - k)^2 =   (W^2+u)z^2-(2\overline{W}+ Wu)z+\left(1+\frac{u^2}{4}\right)\quad.
\end{equation}  
The coefficients of the $z^1$ and $z^0$ terms give two distinct expressions
for $k$.  As they  must be equal, we have
\begin{equation}\label{eqn:linearroot}
  k = \frac{2\overline{W} + Wu}{2 \sqrt{W^2 + u}}
    = {{\sqrt{1 + u^2/4}}} \quad ,
\end{equation}
which gives a cubic equation in $u$,
\begin{equation}
   u^3 + 4(1-W\overline{W})u +4(W^2-\overline{W}^2) = 0 \quad .
\end{equation}

The solution to the cubic can be found from Cardano's formula, with a root
\begin{equation}
   u_1 = \left(\sqrt[3]{2(\overline{W}^2-W^2) +8\sqrt{A}}+\sqrt[3]{2(\overline{W}^2-W^2) -8\sqrt{A}}\right) \quad 
\end{equation}
where
\begin{equation}
A \equiv \frac{1}{27} (1-W\overline{W})^3 + \frac{W^2-\overline{W}^2}{16}\quad .
\end{equation}

Using equations (\ref{eqn:linearsquared}) and (\ref{eqn:linearroot}) our original quartic can be re-written as
\begin{equation}
   \left(z^2-Wz + \frac{u}{2}\right)^2 = (W^2+u)\left(z - \frac{\sqrt{u^2/4+1}}{\sqrt{W^2+u}}\right)^2
\end{equation}

Taking square roots of both sides, we get a quadratic equation for z
\begin{equation}
\left(z^2-Wz + \frac{u}{2}\right) =
\widehat{\pm}\left(z\sqrt{W^2+u} - \sqrt{u^2/4+1}\right) 
\end{equation}
with solution
\begin{eqnarray}
  2e^{i\theta} & = &  W \widehat{\pm} \sqrt{u+W^2} \nonumber \\
  &\widetilde{\pm}& \sqrt{(W \widehat{\pm} \sqrt{u + W^2})^2 - 2\left(u \widehat{\pm} \sqrt{u^2 + 4}\right)} \quad ,
\end{eqnarray}
Note that $\widehat{\pm}$ arises from taking the square root of the
quartic while $\widetilde{\pm}$ arises from the solution of the
resulting quadratic.  We then have four solutions,
$\theta_{\widehat+\widetilde+}$,
$\theta_{\widehat+\widetilde-}$,
$\theta_{\widehat-\widetilde+}$ and
$\theta_{\widehat-\widetilde-}$.


\section{Derivation of the Kassiola \& Kovner configuration Invariant}\label{biquadratic_2_0}

Let $\theta_1, \theta_2, \theta_3, \theta_4$ be the 4 different
solutions of the ACLE measured with respect to the position
angle of the major axis of the potential, $\psi$, which we may
take to be zero without loss of generality,

\begin{equation}\label{general_quartic}
    Ae^{4i\theta} + Be^{3i\theta} + Ce^{2i\theta} + De^{i\theta} + E = 0,
\end{equation}
where $C=0$.  
Applying Vieta's formula to the coefficient of $2^{nd}$ degree in equation (\ref{general_quartic}), we have
\begin{equation}\label{arranged_equation}
    C = e^{i(\theta_1 + \theta_2)} + e^{i(\theta_3 + \theta_4)} + e^{i(\theta_1 + \theta_3)} + e^{i(\theta_2 + \theta_4)} + e^{i(\theta_1 + \theta_4)} + e^{i(\theta_2 + \theta_3)} = 0.
\end{equation}
Using the property,
\begin{equation}
    e^{i\alpha} + e^{i\beta} = 2\cos\left(\frac{\alpha -\beta}{2}\right)e^{i\frac{\alpha + \beta}{2}},\nonumber
\end{equation}
we observe that
\begin{eqnarray}\label{eqn: arranged2}
    \mkern-36mu 2 \cos\frac{\theta_1 + \theta_2 - \theta_3 - \theta_4}{2} e^{i\frac{\theta_1 + \theta_2 + \theta_3 + \theta_4}{2}} + 2\cos\frac{\theta_1 + \theta_3 - \theta_2 - \theta_4}{2} e^{i\frac{\theta_1 + \theta_2 + \theta_3 + \theta_4}{2}} \quad \quad \nonumber \\
+ 2\cos\frac{\theta_1 + \theta_4 - \theta_2 - \theta_3}{2} e^{i\frac{\theta_1 + \theta_2 + \theta_3 + \theta_4}{2}} =0, 
\end{eqnarray}
implying
\begin{equation}\label{eqn: KK inv}
\cos\frac{\theta_1 + \theta_2 - \theta_3 - \theta_4}{2} + \cos\frac{\theta_1 + \theta_3 - \theta_2 - \theta_4}{2} + \cos\frac{\theta_1 + \theta_4 - \theta_2 - \theta_3}{2} =0,
\end{equation}
which is  the \cite{KassiolaKovner} configuration invariant.\\

One can restore full generality using a second relation derived
by Kassiola and Kovner, giving the major axis of the potential,
\begin{equation}
    \psi = \frac{\t_1 + \t_2 + \t_3+ \t_4}{4} \pm \frac{\pi}{4} \quad .
\end{equation}

With similar algebraic gymnastics one can derive the ACLE from
the configuration invariant.

\section{The Witt-Wynne construction for lens-centered shear}\label{app:SIS+XS conics appendix}
\citet{Luhtaru_2021}
expanded the Witt-Wynne construction of
\citet{wynne2018robust} to
the more general case of an SIEP potential with external shear (XS)
aligned with the ellipticity.  Following \citet{Witt_1996}'s initial
development they took the shear to be centered on the source.  We show
here that the Witt-Wynne construction also works for a singular
isothermal sphere (SIS) with lens-centered shear,
\begin{equation}
 \Phi(x,y) = b\sqrt{x^2 +  y^2} - \frac{\gamma}{2}(x^2 - y^2) \quad, 
\end{equation}
giving a potential elongated along the x-axis for positve shear.

Starting with the lens equation,
\begin{equation}
    \mathbf{r} -\mathbf{r}_s = \nabla{\Phi(\mathbf{r})},
\end{equation}
we get
\begin{eqnarray}
    x-x_s & = & \frac{bx}{\sqrt{x^2 +y^2}} - \gamma x \quad {\rm and}\\
    y-y_s & = & \frac{by}{\sqrt{x^2 +y^2}} + \gamma y \quad ,
\end{eqnarray} 
which can be rewritten as
\begin{eqnarray}
    (1+\gamma)x - x_s & = & \frac{bx}{\sqrt{x^2+y^2}} \label{lens_equation_x_component} \quad {\rm and} \\
    (1-\gamma)y - y_s & = & \frac{by}{\sqrt{x^2 + y^2}} \label{lens_equation_y_component} \quad .
\end{eqnarray}
Dividing Equation (\ref{lens_equation_y_component}) by Equation (\ref{lens_equation_x_component}), we get the equation for Witt's hyperbola on which all four images should lie,
\begin{equation}\label{Witt Hyperbola}
    \frac{(1-\gamma)y-y_s}{(1+\gamma)x - x_s} = \frac{y}{x} \quad ,
\end{equation}
which is centered at
\begin{equation}
  (x_h,  y_h) = \left(\frac{x_s}{2\gamma}, -\frac{y_s}{2\gamma}\right) \quad .
\end{equation}    
Squaring Equations (\ref{lens_equation_x_component}) and
(\ref{lens_equation_y_component}) and adding gives us the Wynne ellipse,
\begin{equation}
\left[{x - x_s/(1 + \gamma) \over 1/(1 + \gamma)}\right]^2 +
\left[{y - y_s/(1 - \gamma) \over 1/(1 - \gamma)}\right]^2 = b^2
\quad . 
\end{equation}
It is stretched along the $y$-axis (the short
axis of the potential in our convention) by $1/(1-\gamma)$
and squeezed along the $x$-axis by $1/(1+\gamma)$.
The axis ratio of Wynne's ellipse is therefore
\begin{equation}
  q = \left(\frac{1 - \gamma}{1 + \gamma}\right) \quad . 
\end{equation}
Note that in contrast to the source-centered shear case,
the center of Wynne's ellipse is {\it not} coincident with the source 
but is given instead by
\begin{equation}
   (x_e, y_e) =
   \left(\frac{x_s}{1 + \gamma} ,
        \frac{y_s}{1 - \gamma}\right) \quad .
\end{equation}    
The center of the ellipse lies on the hyperbola, as expected for the
Witt-Wynne construction.  

\section{The ACLE in terms of eccentric angle $\eta$:
algebraic details}\label{app:SIQP_SISXS_equation_relation}
Equation (\ref{hyperbola_ellipse_combined_equation}) in 
Section 6.1 gives Witt's hyperbola recast in terms of the
eccentric angle $\eta$ on the auxiliary circle that corresponds
to a solution of the lens equation on Wynne's ellipse,
\begin{equation}
\label{Acombined}
\frac{\sin{\eta}}{\cos{\eta}} = \left(\frac{1+\gamma}{1-\gamma}\right)\left(\frac{y_s + b\sin{\eta}}{x_s +  b \cos{\eta}}\right) \quad , \\
\end{equation}
where the coordinates of the source are relative to the center of the lens.
 Substituting $\sin{\alpha} = \frac{e^{i\alpha}-e^{-i\alpha}}{2i}$ and $\cos{\alpha} = \frac{e^{i\alpha}+e^{-i\alpha}}{2}$ in Equation (\ref{Acombined}) gives
\newcommand{\e}{e^{i\eta}}
\newcommand{\me}{e^{-i\eta}}
\begin{equation}
    \frac{1}{i}\frac{\e - \me}{\e + \me} = \left(\frac{1+\gamma}{1-\gamma}\right)\frac{2y_s + b\left(\frac{\e-\me}{i}\right)}{2x_s + b (\e+\me)} \quad .
\end{equation}
Cross multiplying we have    
\begin{eqnarray}
    2x_s(1-\gamma) (e^{3i\eta}-e^{i\eta}) + b(1-\gamma)(e^{4i\eta}-1) = 2iy_s(1+\gamma)(e^{3i\eta} + e^{i\eta}) + b(1+\gamma)(e^{4i\eta} -1) \nonumber \\
    (e^{4i\eta} -1)\left(2b\gamma\right) - 2e^{3i\eta}((1-\gamma)x_s - i(1+\gamma)y_s) + 2e^{i\eta}((1-\gamma)x_s + i(1+\gamma)y_s) = 0 \quad , 
\end{eqnarray}
which after gathering terms gives
\begin{equation}
    e^{4i\eta} - 2\left(\frac{(1-\gamma)x_s - i (1+\gamma) y_s}{2b\gamma}\right) e^{3i\eta}  +2\left(\frac{(1-\gamma)x_s + i (1+\gamma) y_s}{2b\gamma}\right) e^{i\eta} - 1 = 0 \quad.
\end{equation}
If we let
\begin{equation}
W = \frac{(1-\gamma)x_s- i(1+\gamma)y_s}{2b\gamma}
\end{equation}
we have
\begin{equation}
    e^{4i\eta} - 2W e^{3i\eta}  + 2\bar{W}e^{i\eta} - 1 = 0 \quad ,
\end{equation}
which is precisely the asymptotically circular lens equation.
The eccentric angles associated with elliptical configurations formed
by SIS+XS are therefore solutions of the ACLE.

\section{Image Magnifications for the SIQP and the SIS+XS potential}\label{app:SISXS_magnification}

\subsection{The singular isothermal quadrupole}
The expression of inverse magnification for each image is given by the determinant of the Jacobian, which \cite{Finch_2002} give as
\begin{equation}
    \mu^{-1}_i = \left(1 - \frac{\partial^2{\Phi}}{\partial x^2}\right)\left(1 - \frac{\partial^2{\Phi}}{\partial y^2}\right) - \left(\frac{\partial^2{\Phi}}{\partial{x}\partial y}\right)^2
\end{equation}
Calculating and substituting the respective partial derivatives, we get
\begin{equation}\label{inverse_mag_eq}
    \mu_{SIQP}^{-1} = 1-\frac{b(1+3\epsilon \cos(2\t))}{r},
\end{equation}
where $(r, \t)$ are the polar coordinates of the image with $x$-axis as the symmetry axis.

We can eliminate $r$ by insisting that the image must form
at a stationary point of the time delay,
\begin{equation}
    t(\vec{r}) = \frac{D}{c}\left(\frac{1}{2}(\vec{r} - \vec{s})^2  -\Phi(\vec{r})\right) \quad , 
\end{equation}
which can be rewritten, upto additive and multiplicative constants, as
\begin{equation}\label{eqn:time_delay}
    \tilde{t}(r, \t)= \frac{1}{2}r^2 - rs \cos(\t-\phi_s)-\Phi(r, \t) \quad .
\end{equation}
The stationarity condition for the time delay at the images is
$ \frac{\partial \tilde{t}}{\partial r} = 0$ and
$\frac{1}{r}\frac{\partial \tilde{t}}{\partial \t} = 0$,\footnote{Note
that the $\frac{1}{r}\frac{\partial \tilde{t}}{\partial \t} = 0$, gives
rise to ACLE for the SIQP \citep{KassiolaKovner}.}
the former of which gives
\begin{equation}     \label{r-phi equation in SIQP}
    r = s\cos(\theta - \phi_s) + b(1-\epsilon \cos 2\theta) 
\end{equation}
for each of the four images.
Substituting $r$ from Equation (\ref{r-phi equation in SIQP}) into
Equation (\ref{inverse_mag_eq}), we get
\begin{equation}
    \mu_{SIQP}^{-1}  = \frac{ s \cos(\t-\phi_s)-4\epsilon b \cos(2\t)}{s\cos(\t - \phi_s) + b(1- \epsilon \cos(2\t))}.
\end{equation}
\subsection{The Singular Isothermal Sphere with External Shear}

The magnifications of the images produced from SIS+XS potential are similarly
found to be
\begin{eqnarray}\label{eqn:inverse_magnification_SISXS_with_r}
    \mu^{-1}_{SIS+XS} = \left(1 - \frac{\tilde{b}y^2}{(x^2 + y^2)^{3/2}} +\gamma \right)\left(1 - \frac{\tilde{b}x^2}{(x^2 + y^2)^{3/2}} - \gamma\right) - \left(- \frac{\tilde{b} x y}{(x^2 + y^2)^{3/2}}\right)^2 \nonumber \\
    = 1-\gamma^2 - \frac{\tilde{b}}{\sqrt{x^2+y^2}} - \frac{\tilde{b} \gamma (x^2-y^2)}{(x^2+y^2)^{3/2}} \quad . \nonumber \\
    = 1-\gamma^2 -
    \frac{\tilde{b}}{r}
    \left[{1+\gamma \cos(2\tilde{\t})}\right]
\end{eqnarray}
Note that we use $\tilde{\t}$ and $\theta$ to distinguish between
the polar angles of images formed by the SIS+XS and SIQP, respectively.

From the stationarity requirement, we have
\begin{equation}
    r - \tilde{s}\cos(\tilde{\t} - \tilde{\phi}_s) - \tilde{b} + \gamma r \cos 2\tilde{\t} = 0 
\end{equation}
and
\begin{equation}
  \frac{1}{r}  =
  \frac
   {1+\gamma \cos2\tilde{\theta}}   
   {\tilde{b} + \tilde{s}\cos(\tilde{\theta} - \tilde{\phi}_s)}
\end{equation}
Substituting $\frac{1}{r}$ into Equation (\ref{eqn:inverse_magnification_SISXS_with_r}),
\begin{eqnarray}\label{eqn:inverse_magnification_SISXS}
    \mu^{-1}_{SIS+XS} = 1-\gamma^2 - \tilde{b}
    \left(
    {\frac
        {1+\gamma \cos2\tilde{\t}}
        {\tilde{b} + \tilde{s}\cos(\tilde{\t} - \tilde{\phi}_s)}
    }
    \right)    
    \left[1+\gamma \cos(2\tilde{\t})\right]
 \nonumber \\
    \mu_{SIS+XS} = \frac{\tilde{b}+\tilde{s}\cos(\tilde{\t}-\tilde{\phi}_s)}{(1-\gamma^2)\tilde{s}\cos(\tilde{\t}-\tilde{\phi}_s)-2\tilde{b}\gamma \cos{2\tilde{\t}}-\tilde{b}\gamma^2 (1+\cos^2{2\tilde{\t}})}
\end{eqnarray}

\vfill
\eject
\bibliography{References}{}
\bibliographystyle{aasjournal}



\end{document}